# LAMA: LAMOST Medium-Resolution Spectral Analysis Pipeline

Chun-qian Li[1,2,3], Jian-rong Shi[1,2], Hong-liang Yan[1,2,3], Zhong-rui Bai[1,2], Jiang-tao Wang[1], and Ming-yi Ding[1,2]  
[1] CAS Key Laboratory of Optical Astronomy, National Astronomical Observatories, Chinese Academy of Sciences, Beijing 100101, People's Republic of China; sjr@nao.cas.cn, hlyan@nao.cas.cn  
[2] School of Astronomy and Space Science, University of Chinese Academy of Sciences, Beijing 100049, People's Republic of China  
[3] Institute for Frontiers in Astronomy and Astrophysics, Beijing Normal University, Beijing 102206, People's Republic of China  
Received 2024 January 19; revised 2024 March 6; accepted 2024 April 1; published 2024 July 15

## Abstract

The Large Sky Area Multi-Object Fiber Spectroscopic Telescope (LAMOST) has obtained more than 23 million spectra, opening an unprecedented opportunity to study stellar physics, as well as the formation and evolution of our Milky Way. In order to obtain the accurate stellar parameters, we develop a LAMOST Medium-Resolution Spectral Analysis Pipeline (LAMA), which estimates the stellar parameters from the LAMOST medium-resolution spectra, including the effective temperature ($T_{\rm eff}$), surface gravity (log $g$), metallicity ([Fe/H]), radial velocity, and rotational velocity ($v \sin i$). LAMA estimates these parameters utilizing the template-matching method. The comparison between our results and those from the high-resolution ones, including APOGEE, GALAH, and PASTEL, shows no obvious bias, indicating the reliability of our results. The accuracy of $T_{\rm eff}$ and [Fe/H] can reach 75 K and 0.12 dex, respectively, for the LAMOST Medium-Resolution Spectroscopic Survey (MRS) spectra with a signal-to-noise ratio higher than 10. For dwarfs, the uncertainty of log $g$ is around 0.17 dex, while, for giants, it ranges from 0.18 to 0.30 dex, with the errors decreasing as log $g$ increases. Using LAMA for the LAMOST-MRS spectra, we estimate the stellar parameters of 497,412 stars. This sample will be very helpful for investigating the formation and evolution of our Galaxy.

*Unified Astronomy Thesaurus concepts:* Catalogs (205); Surveys (1671); Fundamental parameters of stars (555); Spectroscopy (1558)

## 1. Introduction

Large-scale spectroscopic surveys provide us with the opportunity to gain insights into stellar formation in the Galaxy, as well as into stellar dynamic and chemical evolution. On one hand, the stellar parameters and elemental abundances serve as a foundation for studying the stellar formation and evolution. On the other hand, the massive data set produced by the spectroscopic surveys requires automated pipelines to derive the stellar parameters and elemental abundances.

The precise stellar atmospheric parameters are important for determining elemental abundances. The common methods for determining stellar atmospheric parameters are template matching and data driven. Template-matching methods search for the minimum $\chi^2$ value between template and observed spectra to estimate the stellar parameters, while, data-driven methods train the relations between a data set of spectra and the labels of stellar parameters to estimate the stellar parameters. Large spectral sky surveys have developed their own codes or pipelines to analyze spectra, and obtain stellar parameters. For instance, Sloan Extension for Galactic Understanding and Exploration (Yanny et al. 2009) uses SSPP (Lee et al. 2008), the Apache Point Observatory Galactic Evolution Experiment (APOGEE; Majewski et al. 2017) employs ASPCAP (García Pérez et al. 2016), the Third Data Release of the Galactic Archaeology with HERMES (GALAH DR3; De Silva et al. 2015; Buder et al. 2021) utilizes Spectroscopy Made Easy (SME; Valenti & Piskunov 1996; Piskunov & Valenti 2017), Gaia-ESO (Gilmore et al. 2012) applies a multiple pipelines strategy (Smiljanic et al. 2014), the Radial Velocity Experiment (Steinmetz et al. 2006) uses MEDERA (Steinmetz et al. 2020), and Gaia (Gaia Collaboration et al. 2016) utilizes Apsis (Creevey et al. 2023).

The Large Sky Area Multi-Object Fiber Spectroscopic Telescope (LAMOST) is a reflecting Schmidt telescope with a field of view of 5°, and it is able to observe 4000 objects simultaneously with 16 spectrographs and 32 CCD cameras (Cui et al. 2012; Luo et al. 2015). Over the past decade, LAMOST has successively launched the Low-Resolution Spectroscopic (LRS) and Medium-Resolution Spectroscopic (MRS) surveys. To date, LAMOST has taken more than 23 million spectra. This vast collection offers unprecedented opportunities to explore the physics of stars, and the formation and evolution of the Milky Way (Liu et al. 2020; Yan et al. 2022). For analyzing the LAMOST-LRS and MRS spectra, several pipelines using the template-matching or data-driven methods have been developed to obtain the stellar parameters. Template matching includes the following: the LAMOST stellar parameter pipeline (LASP; Wu et al. 2014; Luo et al. 2015) and the LAMOST stellar parameter pipeline at Peking University (LSP3; Xiang et al. 2015). Data-driven methods include the following: KPCA (Xiang et al. 2017), SP_Ace (Boeche et al. 2018), SLAM (Zhang et al. 2020), SPCANet (Wang et al. 2020), and SCDD (Xiang et al. 2021).

In this work, we develop the LAMOST Medium-Resolution Spectral Analysis Pipeline (LAMA)[4] which uses the template-matching method to determine the stellar parameters of F-, G-, and K-type stars. These parameters include effective temperature ($T_{\rm eff}$), surface gravity (log $g$), metallicity ([Fe/H]), radial

---

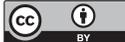

[4] The code can be download from https://github.com/gumplee007/LAMA or doi:10.5281/zenodo.11160083.

velocity (RV), and rotational velocity ($v \sin i$). LAMA adopts the synthetic spectra as the templates and calculates chi-square ($\chi^2$) between the observed and template spectra to derive the stellar parameters.

We give a brief overview of the selection of LAMOST-MRS spectra in Section 2. In Section 3, we describe the methodology of LAMA in detail, and in Section 4, we present our results, including stellar parameters for the LAMOST-MRS spectra, the comparison with other spectroscopic surveys, and calibrated stellar parameters and uncertainties. In the last section, we give a summary.

## 2. Data

Following the initial 5 yr LRS survey, LAMOST launched the MRS survey in 2018. Each MRS spectrum has two arms, i.e., the blue and red ones. The wavelength range spans from 4950 to 5350 Å for the blue arm, and from 6300 to 6800 Å for the red arm. Both arms have a resolution power of approximately $R \sim 7500$. In addition to the stellar atmospheric parameters and RV, the higher resolution of MRS spectrum enables the derivation of stellar rotational velocities, and measurement of the abundance of up to 20 elements (Liu et al. 2020).

After a 1 yr test observation, the LAMOST-MRS began in 2018 September. Up to 2021 June, the LAMOST-MRS has accumulated approximately 30 million spectra, and they are released in the ninth data release (DR9), corresponding to over 1.5 million stars. Figure 1 illustrates the distribution of signal-to-noise ratio (S/N) as a function of Gaia G magnitude for the LAMOST-MRS DR9 spectra. The range of Gaia G magnitude spans from 10 to 15 mag. Generally, the S/N of red arm spectra is higher than that of the blue arm.

### 2.1. Data Selection

It is important to select the spectra before estimating the stellar parameters with LAMA. We select the spectra based on several criteria: the spectral quality, binarity, and the resolution as measured with the thorium–argon (Th–Ar) arc lamps. In the LAMOST-MRS general catalog, the flag *fibermask* indicates fiber problems, while *bad_b* and *bad_r* denote issues with the blue or red arm spectra, respectively,[5] and we only use the spectra with the *fibermask*, *bad_b*, and *bad_r* of 0. S/N is another essential factor in evaluating spectral quality, and we only select those spectra with S/Ns of both the blue and red arms greater than 10.

Multiple systems play an important role in formation and evolution of stars and the Galaxy. Approximately half of objects in the Galaxy are in binary or multiple star systems (Raghavan et al. 2010; Duchêne & Kraus 2013). However, the spectra of multiple star systems, especially those of multiline systems, can lead to inaccurate stellar parameters. Figure 2 shows the spectrum of a spectroscopic binary, where the spectral lines are split due to the RV difference between the binary companions.

By utilizing the cross-correlation function (CCF) and successive derivation (Merle et al. 2017), we identify 5529 double-line and 1610 triple-line candidates from DR9 data set, and they have been excluded before proceeding to the next step.

In order to estimate the stellar parameters of a LAMOST-MRS spectrum, it is necessary to consider the variation in resolution, which fluctuates over time and with fibers. We use the emission lines of Th–AR arc lamp to measure the daily resolution of each spectrum. The detailed method will be described in Section 3.5.

Finally, we select 3,768,587 spectra, corresponding to 594,225 stars, to estimate their stellar parameters with LAMA.

### 2.2. Template Library

There are two different types of spectral libraries used as templates, i.e., empirical and synthetic. The empirical library ELODIE (Prugniel & Soubiran 2001; Prugniel et al. 2007) is used in LASP (Wu et al. 2014; Luo et al. 2015), while the MILES (Sánchez-Blázquez et al. 2006; Falcón-Barroso et al. 2011) is adopted by LSP3 (Xiang et al. 2015). It is noted that the spectroscopic observations lack metal-poor and chemically peculiar stars, posing challenges in estimating the stellar parameters of those stars using empirical libraries. Synthetic libraries, including ATLAS (Kurucz 1979), MARCS (Gustafsson et al. 2008), PHOENIX (Hauschildt & Baron 1999; Husser et al. 2013), among others, offer a wider and more uniform parameter space. They are extensively used for estimating stellar parameters and elemental abundances. Nevertheless, the incomplete line lists and the incorrect atmospheric models of synthetic libraries are the sources of systematic uncertainties. In this work, we utilize the stellar atmospheric models from ATLAS9 to build the templates.

By applying the ATLAS stellar atmospheric models with the new opacity distribution functions (Castelli & Kurucz 2003) and using the solar abundance ratio from Asplund et al. (2009), we calculate the templates that contain approximately 30,000 spectra. A $T_{\rm eff}$ step of 100 K is adopted for the range of 3500–7500 K, and 250 K for the range of 7500–12,000 K. The step for $\log g$ is 0.25 dex. For the metal-poor stars with [Fe/H] $\leqslant -1.0$ dex, a step of 0.2 dex is used, while, for the stars of [Fe/H] $> -1.0$ dex, the step is 0.1 dex. These templates are computed using the SPECTRUM program (Gray & Corbally 1994) with the line list file *lukeall2.iso.lst*.[6] In accordance with the LAMOST-MRS blue and red arm spectra, the wavelength ranges of the templates in the vacuum are of 4950–5350 Å and 6300–6800 Å, respectively. Because the resolution of LAMOST-MRS spectra is not constant ($R \sim 7500$), and varies with time and fiber (as detailed in Section 3.5), the templates are initially generated with a higher resolution of $R \sim 25,000$, then, we decrease their resolution to match that of the observed spectra. Each LAMOST camera has a 4096 × 4096 CCD (Cui et al. 2012), which means each CCD pixel samples about 0.1 Å; we resample the wavelength step to 0.1 Å.

In addition to $T_{\rm eff}$, $\log g$, and [Fe/H], the microturbulence velocity ($\xi_{\rm t}$) also affects the formation of spectral line and is an important parameter for calculating the synthetic spectra. We derive $\xi_{\rm t}$ using empirical relations based on the $T_{\rm eff}$ and $\log g$. The relations have been adopted from the GALAH survey (Buder et al. 2021) for hot dwarfs ($T_{\rm eff} \geqslant 5500$ K, and $\log g \geqslant 3.5$ dex), and from APOGEE (Holtzman et al. 2018; García Pérez et al. 2016) for giants ($\log g < 3.5$ dex). For hot

---

[5] http://www.lamost.org/dr9/v1.1/doc/mr-data-production-description

[6] https://www.appstate.edu/~grayro/spectrum/spectrum.html

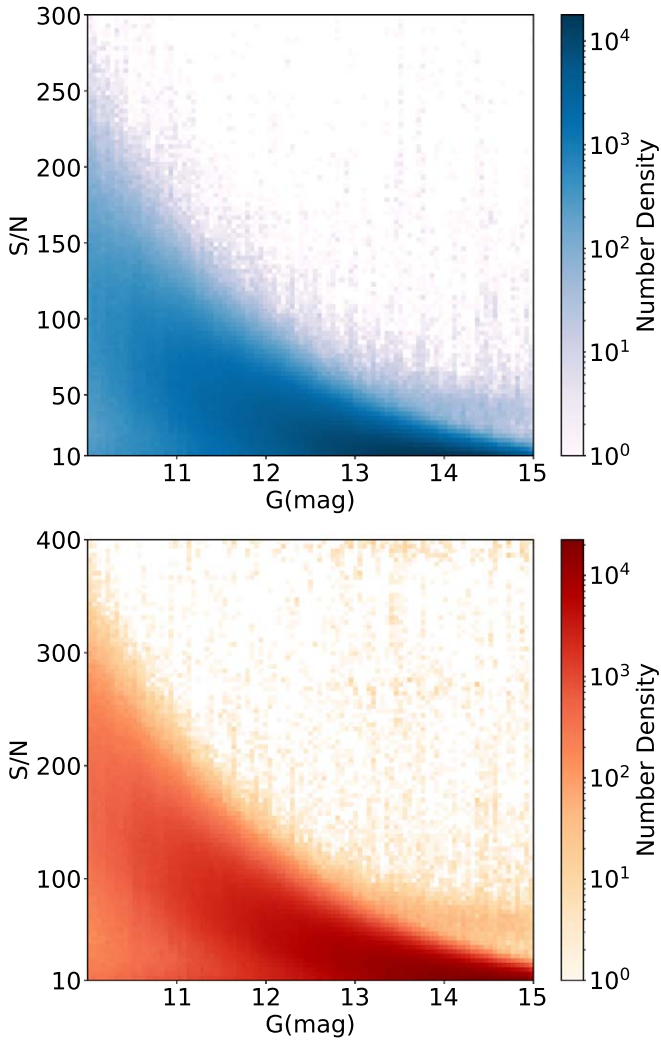

**Figure 1.** The distributions of Gaia G magnitude vs. S/N of the LAMOST-MRS blue (upper panel) and red arm (lower panel) spectra.

dwarfs ($T_{eff} \geqslant 5500$ K), $\xi_t$ is estimated as

$$\xi_t = 1.1 + 1 \times 10^{-4} \cdot (T_{eff} - 5500 \text{ K}) \\ + 4 \times 10^{-7} \cdot (T_{eff} - 5500 \text{ K})^2 \quad (1)$$

while, for cool dwarfs ($T_{eff} < 5500$ K), we set $\xi_t = 1$. For giants, following APOGEE, we determine $\xi_t$ for the stars with [Fe/H] $\geqslant -1.0$ dex and [Fe/H] $< -1.0$ dex using Equations (2) and (3), respectively,

$$\xi_t = 10^{(0.226 - 0.0228 \log g + 0.0297 (\log g)^2 - 0.0113 (\log g)^3)} \quad (2)$$

$$\xi_t = 2.478 - 0.325 \log g. \quad (3)$$

## 3. Methodology

LAMA applies a template-matching method to estimate the stellar parameters, including $T_{eff}$, $\log g$, [Fe/H], RV, and $v \sin i$. In this section, we describe the methods applied in our pipeline. This includes the normalization of spectrum, the measurement of RV, the resolution measurement of LAMOST-MRS, the determination of rotational velocity, and the estimation of the stellar parameters by weighted $\chi^2$.

### 3.1. Flow Chart

Figure 3 represents the flowchart of LAMA. At the beginning, LAMA reads the blue and red arm spectral files. Each pixel in a LAMOST-MRS spectrum is assigned a number, *pixmask*, which is a decimal integer represented by as six-bit binary number. Each bit of this number indicates a specific problem associated with the pixel.[7] Spectra of either the blue or the red arm containing fewer than 3200 data points are discarded. LAMA normalizes the spectrum and then calculates the CCFs between the observed spectra and a series of templates to measure RV. The observed spectrum is then shifted to 0 km s$^{-1}$ (as detailed in Section 3.3). Subsequently, the pipeline calculates $\chi^2$ between the observed spectrum and templates, and estimates the initial stellar atmospheric parameters by weighted $\chi^2$ (Section 3.6). LAMA iterates the processes of normalization and $\chi^2$ minimization once to optimize the spectral normalization (Section 3.2), after which the final stellar parameters are estimated.

### 3.2. Normalization

For the LAMOST-MRS spectra, the flux is uncalibrated and affected by the instrumental response of the telescope and spectrographs. Therefore, a normalization process is essential before measuring RV and estimating the stellar parameters. We first apply a continuum-fitting method to normalize the spectra. After estimating the initial stellar parameters, a continuum-calibration process is performed on the spectra using the nearest template.

The normalization process involves iterative local regression fitting, masking of emission, and absorption lines. Typically, a polynomial fitting is implemented to determine the pseudo-continuum. However, an overfitting issue can arise, particularly when using a high-order polynomial or near the edges of a spectrum. To avoid this issue, our pipeline employs a third-order local polynomial regression fitting (Cleveland 1979; Xu et al. 2019) to normalize both the blue and red spectra of LAMOST-MRS. The local polynomial regression fitting is a nonparametric method for estimating the value of a point by fitting the nearby data points. Comparing with the classical polynomial fitting, this method has the advantage in fitting the local features for a complicated data set.

First, as shown in Figure 4(a), we fit the observed spectra, denoted as $f_1$, using the Python function *localreg*.[8] Through this, we derive the pseudocontinuum $B_1$. During the fitting process, we set the parameter $m_0$ in *localreg* to 135 and 250 Å for the blue and red arm spectra, respectively. These values correspond to one-third and one-half of the wavelength ranges of blue and red arm spectra. The parameter $m_0$ determines the smoothness of the local polynomial regression fitting on the spectra. Additionally, the region of H$_\alpha$ line (6520–6610 Å) in the red arm spectra is masked. After fitting, we calculate the residual ($\epsilon$) between the observed spectrum and the pseudocontinuum, and derive the standard deviation ($\sigma$) of $\epsilon$. LAMA then excludes the spectral data points $i$ with $\epsilon_i > 7\sigma$ corresponding to emission lines or cosmic rays, and the points with $\epsilon_i < 0.5\sigma$ corresponding to absorption lines.

After dividing by the pseudocontinuum, we obtain the first normalized spectrum, denoted as $f_2 = f_1/B_1$. From this spectrum, we select data points without the emission ($\epsilon_i > 7\sigma$) and absorption lines ($\epsilon_i < 0.5\sigma$), denoted as $S_2$. Figure 4(b) illustrates the second normalization step, where a local regression polynomial is used to fit the selected data points $S_2$ to determine the pseudocontinuum, denoted as $B_2$. Following the same process as in the first step, we derive the second normalized spectrum $f_3$. After another iteration of this process, we complete the continuum-fitting, and obtain the final normalized spectrum $f_4$. Generally, cosmic rays manifest as narrow and sharp spikes in the spectrum. In such cases, a sequence of less than 10 emission data points is considered as cosmic rays.

---
[7] http://www.lamost.org/dr9/v2.0/doc/mr-data-production-description
[8] https://github.com/sigvaldm/localreg

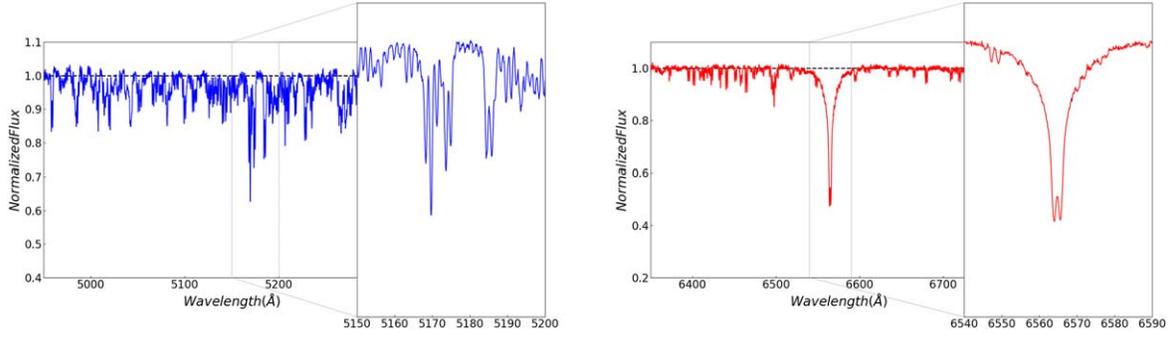

**Figure 2.** The blue and red arm spectra of a double-line spectroscopic binary system. The spectral lines are split due to the RV difference between the companion stars of the binary system.

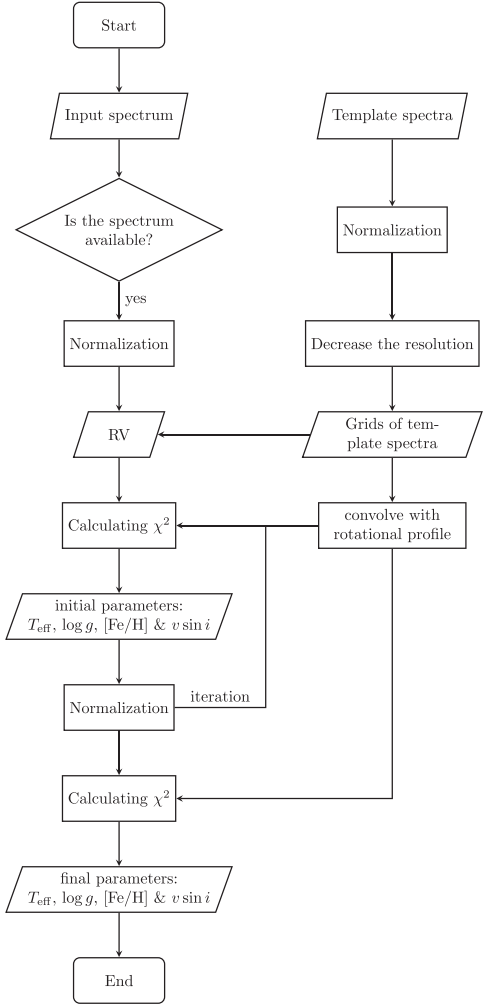

**Figure 3.** Flowchart.

However, the continuum-fitting method is not ideally suited for the LAMOST-MRS spectra. To demonstrate this issue, we calculate a normalized solar-like synthetic spectrum ($T_{\rm eff} = 5800$ K, $\log g = 4.50$ dex, and [Fe/H] = 0.0 dex) matching the LAMOST-MRS wavelength range and resolution. We then add random noise to generate two simulated spectra with S/N of 20 and 100. Subsequently, we determine the pseudocontinuums of these spectra using our method. As shown in Figure 5, the pseudocontinuums of the spectra are obviously different. In the case of the blue arm spectra, the pseudocontinuums for both low and high S/N spectra deviate from the real continuum. For the red arm, although the shape of continuum can be obtained by our method accurately, the pseudocontinuum for the low S/N spectrum is slightly high.

To optimize the normalization, we calibrate the pseudocontinuum of the normalized spectra after estimating the initial stellar parameters. The nearest template is selected from the ATLAS library based on the distance in the parameter space, defined as the following expression:

$$\sqrt{\left(\frac{T_{\rm eff} - T_{\rm eff}'}{100~{\rm K}}\right)^2 + \left(\frac{\log g - \log g'}{0.1~{\rm dex}}\right)^2 + \left(\frac{{\rm [Fe/H]} - {\rm [Fe/H]}'}{0.1~{\rm dex}}\right)^2}; \quad (4)$$

here, $T_{\rm eff}$, $\log g$, and [Fe/H] are the stellar atmospheric parameters of templates, while $T_{\rm eff}'$, $\log g'$ and [Fe/H]$'$ are the initial estimated parameters from the observed spectra. To homogenize the parameter space, we divide the effective temperature, surface gravity, and metallicity by 100 K, 0.1 dex, and 0.1 dex, respectively. As shown in Figure 6, a normalized LAMOST-MRS spectrum using a continuum-fitting method is denoted as $f_4$. The initial stellar parameters of this spectrum are $T_{\rm eff} = 5776$ K, $\log g = 3.12$ dex, and [Fe/H] = $-0.62$ dex, and the nearest parameters of the template are $T_{\rm eff} = 5800$ K, $\log g = 3.00$ dex, and [Fe/H] = $-0.60$ dex. We divide the observed spectrum $f_4$ by the template spectrum denoted as $T$, and obtain $S_4 = f_4/T$. Subsequently, a third-order local polynomial regression is applied to fit the quotient, resulting in the fitted curve $B_4$. The strong line H$_\alpha$ is masked, and the parameter $m_0$ in *localreg* is set as 135 and 250 Å for the blue and red arm quotients, respectively. Finally, by dividing the LAMOST-MRS spectrum with $B_4$, we obtain the calibrated normalized spectrum $f_5 = f_4/B_4$. Then, we iterate the processes of calculating the $\chi^2$, estimating the stellar parameters and calibrating the pseudocontinuum once more.

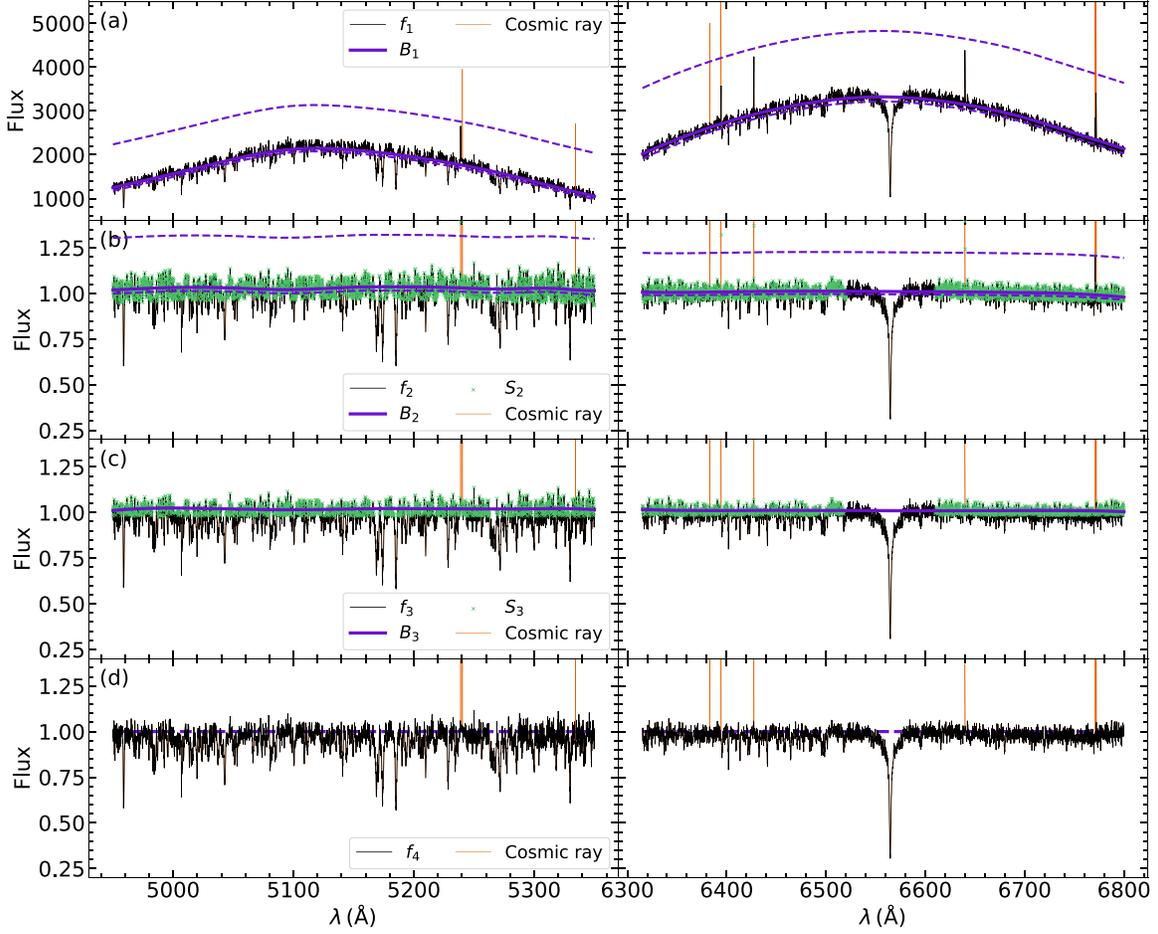

**Figure 4.** The normalization process using the continuum-fitting method. LAMOST-MRS spectra are drawn with the solid black lines; the left panels show the blue arm spectra, while the right panels display the red arm ones. The solid purple lines represent the pseudocontinuums, and the dashed purple lines indicate the upper and lower limits, which are used to exclude the emission and absorption lines. The identified cosmic rays are drawn with the orange lines. (a) The first fitting process, the initial continuum is fitted. (b) The second fitting process, the LAMOST-MRS spectrum is divided by the pseudocontinuum, and the green crosses denote the data points selected for local polynomial regression fitting. (c) The third fitting process, the pseudocontinuums are further refined. (d) The final normalized spectrum with the cosmic rays excluded.

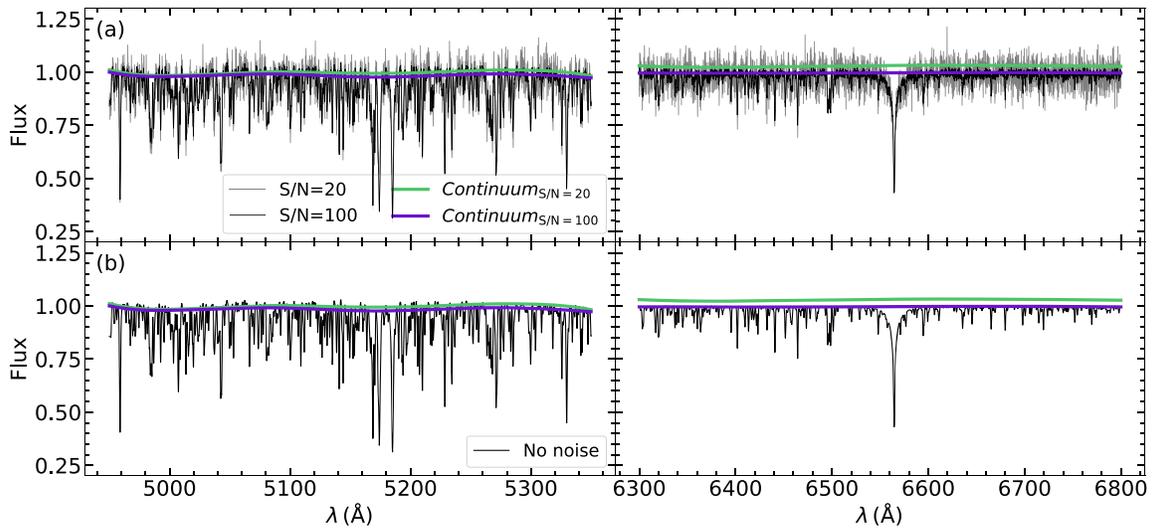

**Figure 5.** The normalized synthetic spectra with different S/N. The stellar parameters of these synthetic spectra are $T_{\rm eff} = 5800$ K, $\log g = 4.50$ dex, and [Fe/H] = 0.0 dex, and the wavelength range and resolution are the same as LAMOST-MRS. (a) The gray and black lines are the spectra with S/N = 20 and 100, respectively. The green and purple lines indicate the pseudocontinuums of these spectra, which are normalized by the continuum-fitting method. (b) The green and purple lines are the same as in panel (a), representing the pseudocontinuums. The black lines are the synthetic spectra without noise.

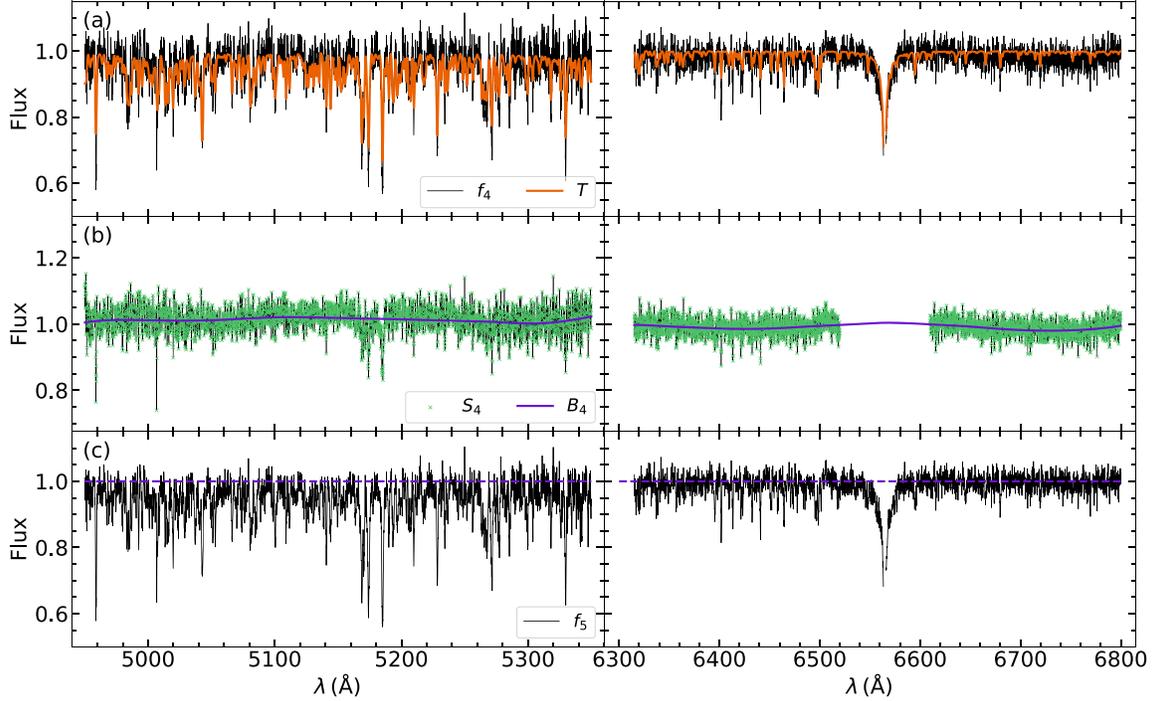

**Figure 6.** The continuum-calibration process. (a) The black line represents the normalized spectrum, and the orange line corresponds the template spectrum, which is the nearest in the parameter space. The stellar parameters of this template spectra are $T_{\rm eff}$ = 5800 K, $\log g$ = 3.00 dex, and [Fe/H] = $-0.60$ dex. (b) The green crosses denote the quotient data points obtained by dividing the observed spectrum by the template one. The purple line is the curve fitted using local polynomial regression. (c) By dividing the normalized spectrum by the fitting curve, we calibrate the normalized continuum of the LAMOST-MRS spectrum.

### 3.3. Radial Velocity

To measure the RVs of LAMOST-MRS blue and red arm spectra, we adopt CCF. The CCF value for a given RV is calculated using the normalized discrete CCF, defined as

$$CCF(v) = \sum_{i=1}^{n} \frac{(O_i - \overline{O})}{\sigma_O} \cdot \frac{(T_{i,v} - \overline{T})}{\sigma_T} \quad (5)$$

where $v$ is a given RV value, $O_i$ and $T_{i,v}$ are the flux values of observed and template spectra at $i$th wavelength, respectively, and the wavelength of template is shifted by $v$. $\overline{O}$ and $\sigma_O$ are the mean and standard deviation of the flux values of the observed spectrum, while $\overline{T}$ and $\sigma_T$ represent the same for the template. The RV template grid contains 15 spectra, covering a range of $T_{\rm eff}$ from 4000 to 8000 K, in steps of 1000 K. We use the templates with solar metallicity ([Fe/H] = 0 dex) and three different $\log g$ values: 1.0, 3.0, and 5.0 dex. After calculating the 15 CCFs between the observed spectrum and the 15 templates, we select the template that yields the maximum CCF amplitude to derive the RV of the spectrum.

In addition, we exclude the multiline spectra and select only the single-line spectra. We determine the maximum value of CCF, representing the RV value of the spectrum. To estimate the RV uncertainties, we implement the Monte Carlo simulation technique described by Li et al. (2021). Once the RVs for both blue and red arms have been measured, we shift the wavelength of the spectra using the formula $\Delta\lambda \equiv \lambda_0 \cdot v/c$, where $\lambda_0$ is the original wavelength, $v$ is the RV value, and $c$ is the speed of light.

### 3.4. Resolution

The resolution of LAMOST-MRS is not a constant value of 7500; rather, it varies over time and with fibers. Due to the variation in the instrumental broadening for each fiber and each observation, it is essential to measure the resolution of each LAMOST spectrum before estimating the stellar parameters.

To determine the resolution of LAMOST-MRS blue and red arm spectra, we measure the full width at half-maximum (FWHM) of the emission lines in the calibration lamp spectra. LAMOST uses Th–Ar calibration arc lamps. We carefully select a set of lines from the blue and red arm lamp spectra, ensuring that they are neither blended nor too strong or too weak. The resolution at the wavelength of each line is calculated using the equation $R \equiv \lambda/\Delta\lambda_{\rm FWHM}$, where $\Delta\lambda_{\rm FWHM}$ is the FWHM of the emission line at wavelength $\lambda$. Considering that the wavelength coverages of both LAMOST-MRS blue and red arm spectra are not wide, we assume the resolution to be constant across each arm. Then, we compute the resolution of each arc lamp spectrum with the emission lines by Tukey's biweight function (Beaton & Tukey 1974).

LAMOST captures arc lamp spectra at the beginning, middle, and end of a night. However, not every observed spectrum is immediately followed by an arc lamp observation. Therefore, we calculate the mean resolution for each fiber over the course of an observed night. This approach allows us to obtain the LAMOST-MRS resolution for each night. The distribution of resolutions for the blue and red arm spectra is illustrated in Figure 7. From this figure, it is evident that the peak resolution of LAMOST-MRS exceeds 7500, and the scatter of the resolution is larger for the blue arm than that for the red arm. Figure 8 shows the variations of resolution with different fibers and over time.

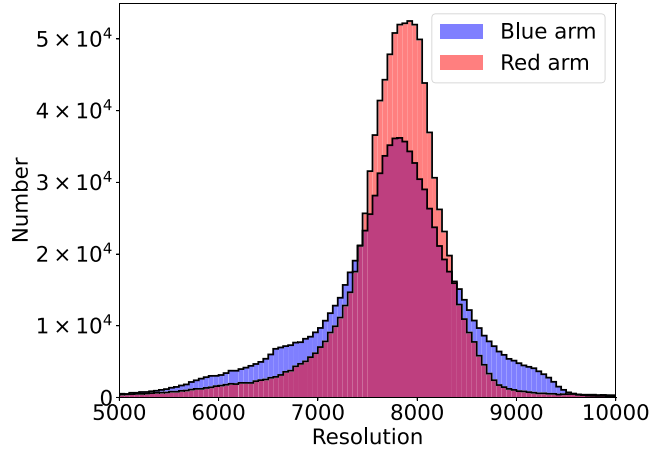

**Figure 7.** The distribution of resolution of the LAMOST-MRS blue and red arm spectra. The peak values of resolution for both the blue and red arm spectra are higher than 7500. Additionally, the scatter of resolution is larger for the blue arm spectra compared to the red spectra.

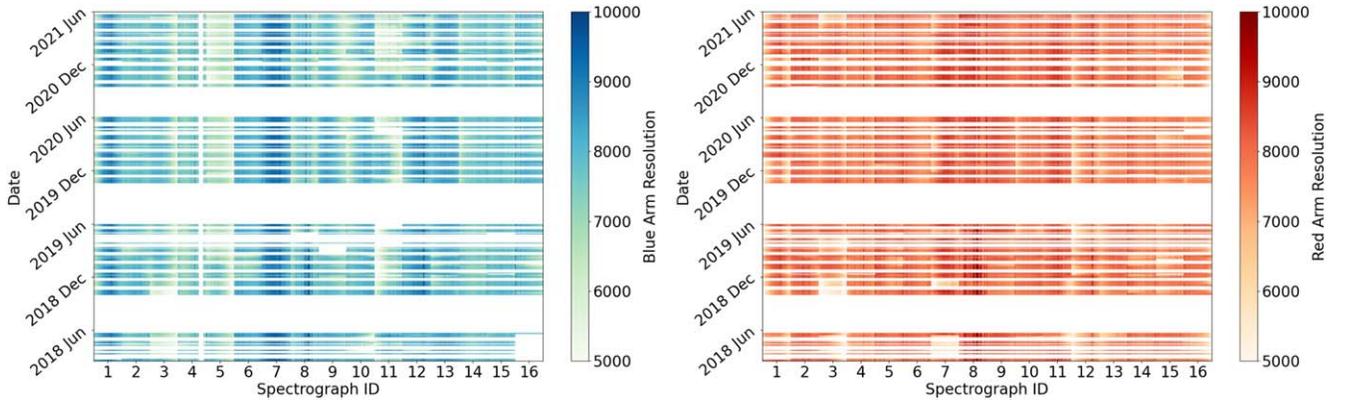

**Figure 8.** The LAMOST-MRS resolutions of 4000 fibers of each night. The resolutions vary with dates and fibers.

### 3.5. Rotation

Utilizing the measured resolution, we are able to derive the rotational velocities from LAMOST-MRS spectra exceeding 40 km s$^{-1}$ ($R \sim 7500$). To estimate the rotational velocity, we convolve the templates with a series of rotational velocity profiles. Subsequently, we calculate the $\chi^2$ values between the observed and the convolved template spectra.

When determining the rotation velocity, it is necessary to consider other external broadening mechanisms, such as instrumental broadening and macroturbulence velocity ($\zeta_t$). We reduce the resolution of templates from 25,000 to match that of observed spectrum by convolving them with a Gaussian kernel. The value of $\zeta_t$ depends on $T_{\rm eff}$ and $\log g$. For dwarfs ($\log g > 3.5$ dex), we calculate $\zeta_t$ using the relation provided by Gray (1984),

$$\zeta_t = \begin{cases} 3.95 \cdot T_{\rm eff}/1000 - 19.25, & T_{\rm eff} > 4873 \text{ K} \\ 0, & \text{else} \end{cases}, \quad (6)$$

and for giants ($\log g < 3.5$ dex), we use a different relation (Gray 1982),

$$\zeta_t = \begin{cases} 7 - (5.5 - T_{\rm eff}/1000)^{2.6}, & T_{\rm eff} < 5500 \text{ K} \\ 7 + (T_{\rm eff}/1000 - 5.5)^{2.6}, & T_{\rm eff} \geqslant 5500 \text{ K} \end{cases}. \quad (7)$$

Then, we convolve the template with a rotational profile as described by Gray (1992),

$$G(\Delta\lambda) = C_1 \sqrt{1 - \left(\frac{\Delta\lambda}{\Delta\lambda_L}\right)^2} + C_2 \left[1 - \left(\frac{\Delta\lambda}{\Delta\lambda_L}\right)^2\right] \quad (8)$$

where $\Delta\lambda_L = \lambda \cdot v \sin i / c$ is the maximum wavelength shift, $\Delta\lambda$ is the wavelength shift caused by stellar rotation, and $\Delta\lambda \leqslant \Delta\lambda_L$. The coefficients $C_1$ and $C_2$ are

$$C_1 = \frac{2(1-\varepsilon)}{\pi\Delta\lambda_L(1-\varepsilon/3)} \quad (9)$$

$$C_2 = \frac{\varepsilon}{2\Delta\lambda_L(1-\varepsilon/3)}; \quad (10)$$

here, $\varepsilon$ is the limb-darkening coefficient, which we set as $\varepsilon = 0.6$. This approach allows us to derive the rotational profile for a given rotational velocity. It is essential to note that the stellar rotation and the macroturbulence contribute to broaden the spectral lines, since their broadening profiles are similar. The maximum wavelength shift due to the above two process is expressed as $\Delta\lambda_L = \lambda \cdot v_b / c$, where $v_b$ is calculated as $v_b = \sqrt{v^2 \sin^2 i + \zeta_t^2}$.

### 3.6. Stellar Parameters Estimated by Weighted $\chi^2$

LAMA estimates the stellar parameters by calculating the weighted $\chi^2$ between the LAMOST-MRS spectrum and a grid of templates. As the wavelength sampling rate of LAMOST-MRS is not constant, we interpolate the observed spectra to 0.1 Å after normalization and RV shift. The $\chi^2$ is then calculated between the observed spectrum and a set of templates using the following formula (Equation (11))

$$\chi^2 = \sum_{i=1}^{n} \frac{(O_i - T_i)^2}{T_i}. \qquad (11)$$

In this equation, for each data point $i$, $O_i$ and $T_i$ represent the flux values of the observed and template spectra at the same wavelength, respectively. It needs to be pointed out that we exclude the spectral region within the FWHM of $H_\alpha$ before calculating $\chi^2$, as the center of this line is formed in the chromosphere, which can affect the accuracy of the parameter estimation.

We adopt the weighted average algorithm method to estimate stellar parameters as described in LSP3 (Xiang et al. 2015). This process selects a subset of the templates for which $\chi^2 < (1 + a)\chi^2_{\min}$, where $\chi^2_{\min}$ is the minimum value of $\chi^2$. $a$ is a free parameter greater than zero. For the LAMOST-MRS spectra, we assign $a = 0.3$ when estimating $T_{\rm eff}$, $\log g$, and [Fe/H], and $a = 1.0$ for $v \sin i$. The weight is determined by the equation

$$w_j \equiv 1 - f \times \frac{\chi^2_j - \chi^2_{\min}}{\chi^2_{\max} - \chi^2_{\min}}; \qquad (12)$$

here, $\chi^2_{\max}$ represents the maximum value of $\chi^2$ within the selected subset. We adopt the fudge factor $f$ as 0.8. This means that the weight value $w = 1.0$ when $\chi^2 = \chi^2_{\min}$, and $w = 0.2$ when $\chi^2 = \chi^2_{\max}$.

Finally, we estimate the stellar parameters using the equation

$$\theta = \frac{\sum_j (w_j \times \theta_j)}{\sum_j w_j}. \qquad (13)$$

Here, $\theta_j$ denotes the stellar parameter of the $j$th template. Therefore, we can estimate the stellar parameters of $T_{\rm eff}$, $\log g$, [Fe/H], and $v \sin i$ for each LAMOST-MRS spectrum.

Since the stellar parameters are sensitive to different wavelength ranges, we select specific arm spectra for estimating them. For $T_{\rm eff}$, when $T_{\rm eff} < 8000$ K, the strength of the Balmer lines in the red arm spectra is sensitive to $T_{\rm eff}$, while it does not depend on surface gravity (Gray 1992). Consequently, we use the red arm spectra for estimating $T_{\rm eff}$. For $\log g$, the Mg Ib triple lines in the blue arm spectra display strong pressure-broadened wings for F-, G-, and K-type stars, making them suitable for $\log g$ estimation. In the case of [Fe/H] and $v \sin i$, the blue arm spectra contain more lines compared to the red ones, making them informative for [Fe/H] and $v \sin i$. However, a reliable $T_{\rm eff}$ is crucial for accurately determining the other parameters. Therefore, it is necessary to use both the blue and red arm spectra for deriving $\log g$, [Fe/H], and $v \sin i$. We define a combined $\chi^2$ as $\chi^2_{\rm combine} = \chi^2_B \times \chi^2_R$. Considering that the red arm spectra generally have higher S/Ns than those of the blue spectra, the $\chi^2$ values of the red arm are generally smaller; we opt to multiply the $\chi^2$ values of each arm instead of adding them.

To reduce the computational time for estimating the stellar parameters, we use subsets of the template library with varying ranges and steps for the stellar parameters. First, for estimating the initial stellar parameters, we set broader steps for templates: 500 K for $T_{\rm eff}$, 1.0 dex for $\log g$ and 0.5 dex for [Fe/H]. Subsequently, as we proceed to the second and final estimation, we gradually narrow down the step size of these parameters in the templates. This refinement allows for more precise estimation as the process advances. Additionally, for each stage of estimation, we select a subset of templates where $\chi^2 < (1 + a)\chi^2_{\min}$. The parameter ranges of this selected subset are used to determine the template pool for the subsequent $\chi^2$ calculations.

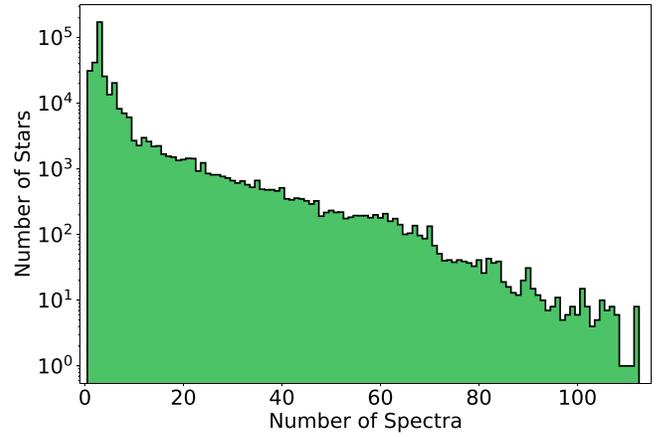

**Figure 9.** The number of stars vs. the number of visits.

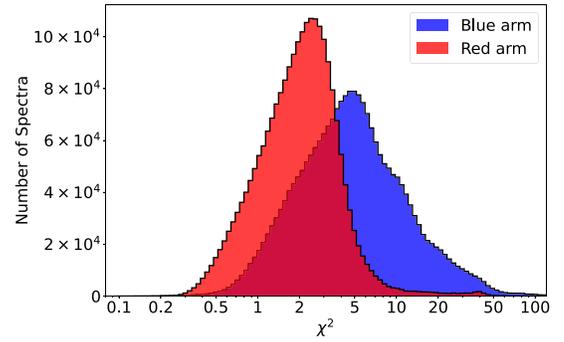

**Figure 10.** The distribution of $\chi^2$ for the blue and red arm spectra.

## 4. Result

In our pipeline LAMA, we estimate the stellar parameters from the LAMOST-MRS DR9 data sets. After excluding the outliers (Section 4.1), performing calibration (Section 4.3), and estimating the uncertainties (Section 4.4), we finally obtain the stellar parameters (available via doi:10.5281/zenodo.11160083), including $T_{\rm eff}$, $\log g$, [Fe/H], RV, and $v \sin i$, along with their uncertainties for 3,126,343 spectra, corresponding to 497,412 FGK stars. Figure 9 displays the number of stars versus the number of visits.

### 4.1. Stellar Parameters of Stars

Following the estimation of the stellar parameters from the LAMOST-MRS spectra, we apply data selection to remove outliers and a weighted average algorithm to derive the stellar parameters of stars with multiple visits.

Outliers are removed based on the $\chi^2$ values, RV, and the presence of the emission $H_\alpha$ line. We calculate the $\chi^2$ between the nearest template and the blue and red arm spectra, denoted as $\chi_B^2$ and $\chi_R^2$, respectively. Figure 10 shows the distribution of these $\chi^2$ values. Spectra with $\chi_B^2 > 40$ or $\chi_R^2 > 10$ are excluded. Additionally, spectra with an RV difference between the blue and red arms exceeding 10 km s$^{-1}$ ($|RV_B - RV_R| > 10$ km s$^{-1}$), and those with RV uncertainties greater than 10 km s$^{-1}$ in blue ($\sigma_{RV,B} > 10$ km s$^{-1}$) or 5 km s$^{-1}$ in the red arm ($\sigma_{RV,R} > 5$ km s$^{-1}$) are rejected. Finally, spectra exhibiting an emission H$_\alpha$ line are omitted.

We calculate the stellar parameters of stars with multiple visits using Equation (13). In the equation, $\theta_j$ represents the stellar parameter of each spectrum of a star, $w_j = 1/\chi_j^2$ is the weight, and $\chi_j^2$ is calculated between the observed spectrum of each visit and its nearest template.

### 4.2. Comparison with Reference Results

To evaluate the accuracy of our stellar parameters, we compare our results with the reference data from high-resolution spectroscopic sky survey APOGEE and GALAH, and the bibliographical compilation catalog PASTEL (Soubiran et al. 2010, 2016).

APOGEE is a high-resolution ($R \sim 22,500$), near-infrared (1.5–1.7 $\mu$m) stellar spectroscopic survey (Majewski et al. 2017). Up to 2021 January, the spectra and the stellar parameters of about 657,000 stars have been released in the 17th data release of the Sloan Digital Sky Survey (Abdurro'uf et al. 2022). The ASPCAP (García Pérez et al. 2016) determines the stellar atmospheric parameters and chemical abundances using $\chi^2$ minimization between observed spectra and a theoretical spectral library.

GALAH, another high-resolution ($R \sim 28,000$) spectroscopic survey, is carried out with a fiber-fed, multiobject spectrograph on the Anglo-Australian Telescope (De Silva et al. 2015). The spectrographs has four arms, blue (4718–4903 Å), green (5649–5873 Å), red (6481–6739 Å), and IR (7590–7890 Å). The third data release of GALAH contains 588,571 nearby stars (Buder et al. 2021), with the stellar parameters provided by the spectrum synthesis code SME (Valenti & Piskunov 1996; Piskunov & Valenti 2017).

The PASTEL catalog compiles the stellar parameters determined from high-resolution and high S/N spectra. The 2016 version of the PASTEL catalog (Soubiran et al. 2016) includes 31,401 stars with the stellar parameters $T_{\rm eff}$, log $g$, and [Fe/H], corresponding to 1142 bibliographical references.

We perform a data selection for APOGEE and GALAH data set before comparison. For APOGEE, the stellar parameter catalog employs the *ASPCAPFLAG* bitmask to indicate potential issues with the ASPCAP fits.[9] We exclude the objects with potential inaccuracies in $T_{\rm eff}$, log $g$, [Fe/H], $\xi_t$, and $v \sin i$, as well as those with warning on these parameters. Additionally, we remove the stars flagged for observational or data reduction problems, as indicated by the 31st to the 41st bit of the *ASPCAPFLAG*. For the GALAH, we select the stars with *flag_sp* = 0 and *flag_fe_h* = 0.[10] In the PASTEL catalog, we calculate the mean value and scatter of $T_{\rm eff}$, log $g$, [Fe/H] for the stars with multiple results. Finally, we select 25,977, 10,299, and 376 common stars from APOGEE, GALAH, and PASTEL, respectively, within the LAMA effective temperature range of 4000 K < $T_{\rm eff}$ < 7000 K.

---

[9] https://www.sdss4.org/dr17/irspec/apogee-bitmasks/
[10] https://www.galah-survey.org/dr3/flags/

**Table 1**
The Coefficients of the Cubic Polynomials

| Parameter | $a_0$ | $a_1$ | $a_2$ | $a_3$ |
|---|---|---|---|---|
| $T_{\rm eff}$ (K) | −1.566(−8) | 2.781(−4) | −1.615(0) | 3.074(+3) |
| log $g$ (dex) | −9.023(−11) | 1.533(−6) | −8.432(−3) | 1.499(+1) |
| [Fe/H] (dex) | −7.218(−11) | 1.159(−6) | −6.057(−3) | 1.031(+1) |

**Note.** The numbers in the round brackets represent the powers of 10. For instance, −1.566(−8) means −1.566 × 10$^{-8}$.

The comparisons of these stellar parameters are depicted in Figures B1, B2, and B3, respectively, and a negligible bias between LAMA and APOGEE or GALAH can be seen. However, the $T_{\rm eff}$ of LAMA is slightly lower compared to PASTEL. For hot stars ($T_{\rm eff} > 6500$ K), the $T_{\rm eff}$ of APOGEE is lower than those of LAMA, and the log $g$ values exhibit a larger scatter for the common stars. Additionally, the log $g$ values for dwarfs from LAMA are slightly lower than all the reference results. In terms of [Fe/H], the results of LAMA show good agreement with APOGEE, GALAH, and PASTEL, even for the metal-poor stars. Due to the resolution of 7500 for LAMOST-MRS spectra, the $v \sin i$ are most consistent with those from APOGEE or GALAH for the objects of $v \sin i > 40$ km s$^{-1}$. The scatters of the differences between LAMA stellar parameters and those from the references depend on both the uncertainties of LAMA and references. The size of the scatters will be increased if the uncertainties of the references are considered.

### 4.3. Calibration

Considering that the wavelength coverage of LAMOST-MRS and GALAH are both visible, and GALAH has higher resolution, we select the common stars between LAMA and GALAH for calibrating the stellar parameters. It should be noted that $T_{\rm eff}$ significantly influences the other parameters; we consider the bias related to $T_{\rm eff}$ during the calibration of other parameters. The difference in stellar parameters between LAMA and GALAH is defined as

$$\Delta \theta \equiv \theta_{\rm GALAH} - \theta_{\rm LAMA}; \quad (14)$$

here, $\theta$ represents a stellar parameter, specifically for $T_{\rm eff}$, log $g$, or [Fe/H]. We select only those LAMA stellar parameters estimated from the LAMOST spectra with S/N > 50.

Since the stellar parameters derived from spectra are not uniformly distributed across $T_{\rm eff}$, we group the parameters in different $T_{\rm eff}$ ranges. The mean value of $T_{\rm eff}$ in the $i$th group is denoted as $\overline{T_{{\rm eff},i}}$. We then calculate the weighted average of the parameter differences $\Delta \theta$ in each group using the following equation:

$$\overline{\Delta \theta_i} \equiv \frac{\sum_j \Delta \theta_j \times \omega_j}{\sum_j \omega_j}; \quad (15)$$

here, $j$ represents the $j$th parameter point in the $i$th group, and the weights $\omega$ are assigned as $1/\chi^2$. We use $\chi_R^2$ for fitting $\Delta T_{\rm eff}$, and $\chi_{\rm combine}^2$ for $\Delta \log g$ and $\Delta$[Fe/H]. Assuming that the differences of stellar parameters are only related to $T_{\rm eff}$, we use a cubic polynomial to fit the bias,

$$\overline{\Delta \theta} = a_0 \cdot \overline{T_{\rm eff}}^3 + a_1 \cdot \overline{T_{\rm eff}}^2 + a_2 \cdot \overline{T_{\rm eff}} + a_3. \quad (16)$$

The coefficients for calibrating the LAMA stellar parameters are listed in Table 1. Figure 11 shows the differences in parameters

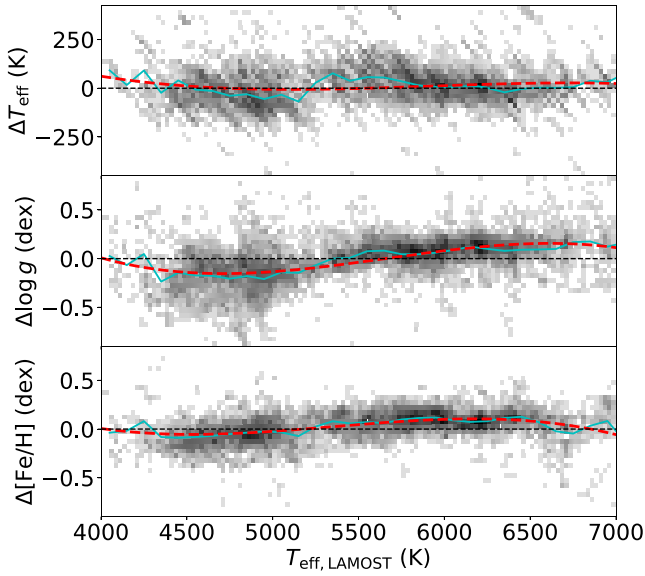

**Figure 11.** The differences in parameters between GALAH and LAMA as a function of $T_{\rm eff}$. The cyan lines represent the average differences for each group of data. The red dashed lines indicate the fitted cubic polynomial curves.

**Table 2**
The Coefficients of the Quadratic Polynomial Equations

| Parameter | $a_0$ | $a_1$ | $a_2$ |
| | $a_3$ | $a_4$ | $a_5$ |
| --- | --- | --- | --- |
| $T_{\rm eff}$ (K) | 4.899(–6) | –7.281(–2) | 3.209(+2) |
| $\log g$ (dex) | 2.509(–8) | 5.585(–2) | –1.051(–4) |
| | 1.599(–4) | 9.201(–2) | –4.852(–1) |
| [Fe/H] (dex) | 4.372(–8) | –6.052(–3) | –1.281(–5) |
| | –4.907(–4) | 3.906(–2) | 1.443(0) |

**Note.** The numbers in the round brackets represent the powers of 10. For instance, 4.899(–6) means $4.899 \times 10^{-6}$.

between LAMA and GALAH as a function of $T_{\rm eff}$, which indicates no significant bias in $T_{\rm eff}$. However, the biases of the $\log g$ and [Fe/H] vary with $T_{\rm eff}$; for hot stars, $\log g$ and [Fe/H] of GALAH are higher than those of LAMA; vice versa for the cool stars.

Afterwards, the calibrated parameter is given by $\theta_{\rm calib} = \theta_{\rm LAMA} + \overline{\Delta\theta}$. Figure A1 shows the distributions of these calibrated stellar parameters. After the calibration, as shown as in Figures B4, B5, and B6 the scatter in differences of the stellar parameters between LAMA and references, including APOGEE, GALAH, and PASTEL, is reduced. Since the calibration of $T_{\rm eff}$ is minor, the improvement is not significant.

### 4.4. Uncertainty Estimation

The precision of stellar parameters is subjected to various stages, starting from observation and extending through spectral reduction and analysis. To estimate the systematic uncertainties of LAMA stellar parameters, we use the stars that are common with other high-resolution surveys. The random uncertainties are derived from stars with multiple visits. After the calibration, the total uncertainty for each stellar parameter ($T_{\rm eff}$, $\log g$, and [Fe/H]) is calculated using the equation

$$\sigma_\theta = \sqrt{\sigma_{\theta,{\rm sys}}^2 + \sigma_{\theta,{\rm rand}}^2} \qquad (17)$$

where $\theta$ is the stellar parameter, and $\sigma_{\theta,{\rm sys}}$ and $\sigma_{\theta,{\rm rand}}$ represent the systematic and random uncertainties. It should be noted that we only estimate the random uncertainty for $v \sin i$.

#### 4.4.1. Systematic Uncertainties

The systematic uncertainties in stellar parameters are primarily induced by the theoretical templates and method. Jofré et al. (2019) have discussed the uncertainty of stellar abundances caused by theory and spectral line list. In this work, we focus on the errors arising from the methods, mainly including normalization and $\chi_2$ minimization.

Normalization of LAMOST-MRS spectra is crucial for measuring RV and calculating $\chi^2$ value. However, there is no universal normalization method applicable for all the stellar parameter spaces. In particular, for cool stars with molecular bands, and spectra with low S/N or blended spectral lines, it is challenging to normalize accurately, making it a significant source of error in estimating the stellar parameters. Another error resource is the method of $\chi^2$ minimization. The number of the templates, the value of $a$ in computing weights (Equation (12)), and the step of iterations used to determine the stellar parameters all have an impact on the errors.

However, it is very challenging to quantify the errors caused by the templates, and the methods of normalization and $\chi^2$ minimization, and it is beyond the scope of this work. To estimate the systematic uncertainties of LAMA stellar parameters, we compare the results of LAMA with those from GALAH. We choose the results of GALAH as the standard stellar parameters because of its sufficient number of high-resolution spectra, providing comprehensive parameter space coverage for stars common to the LAMOST-MRS data set. To minimize the impact of the random errors introduced by noise, we select only the LAMA results derived from the spectra with S/N > 50.

We define the absolute difference of stellar parameters between LAMA and GALAH as

$$\Delta\theta \equiv |\theta_{\rm LAMA} - \theta_{\rm GALAH}|; \qquad (18)$$

and we consider that the systematic uncertainties of stellar atmospheric parameter $T_{\rm eff}$, $\log g$, and [Fe/H] are affected by $T_{\rm eff}$ and the parameters themselves. Thus, we fit the absolute differences with following quadratic polynomial equations:

$$\Delta T_{\rm eff} = a_0 T_{\rm eff}^2 + a_1 T_{\rm eff} + a_2 \qquad (19)$$

$$\Delta \log g = a_0 T_{\rm eff}^2 + a_1 \log g^2 + a_2 T_{\rm eff} \cdot \log g$$
$$+ a_3 T_{\rm eff} + a_4 \log g + a_5 \qquad (20)$$

$$\Delta [{\rm Fe/H}] = a_0 T_{\rm eff}^2 + a_1 [{\rm Fe/H}]^2 + a_2 T_{\rm eff} \cdot [{\rm Fe/H}]$$
$$+ a_3 T_{\rm eff} + a_4 [{\rm Fe/H}] + a_5. \qquad (21)$$

As the parameter calibration process described in Section 4.3, we group the stellar parameters based on the $T_{\rm eff}$, $\log g$, or [Fe/H] values. Within each group, we calculate the weighted average of the absolute differences. Subsequently, we derive the systematic uncertainties with Equations (19)–(21), and the coefficients obtained from the fitting are listed in Table 2.

#### 4.4.2. Random Uncertainties

The random uncertainties of stellar parameters stem from instruments, for example, the telescopes, spectrographs, and detectors. We select the stars with multiple visits to estimate the

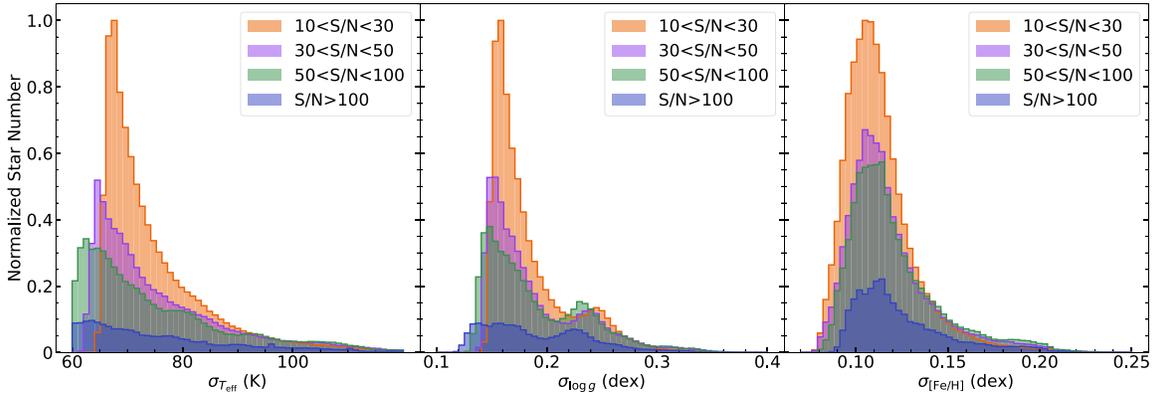

**Figure 12.** The distribution of uncertainties for the stellar parameters $T_{\rm eff}$, $\log g$, and [Fe/H]. The S/Ns of spectral are drawn with different colors. The orange histogram represents the uncertainties of spectra with $10 < {\rm S/N} < 30$, purple for $30 < {\rm S/N} < 50$, green for $50 < {\rm S/N} < 100$, while blue for ${\rm S/N} > 100$.

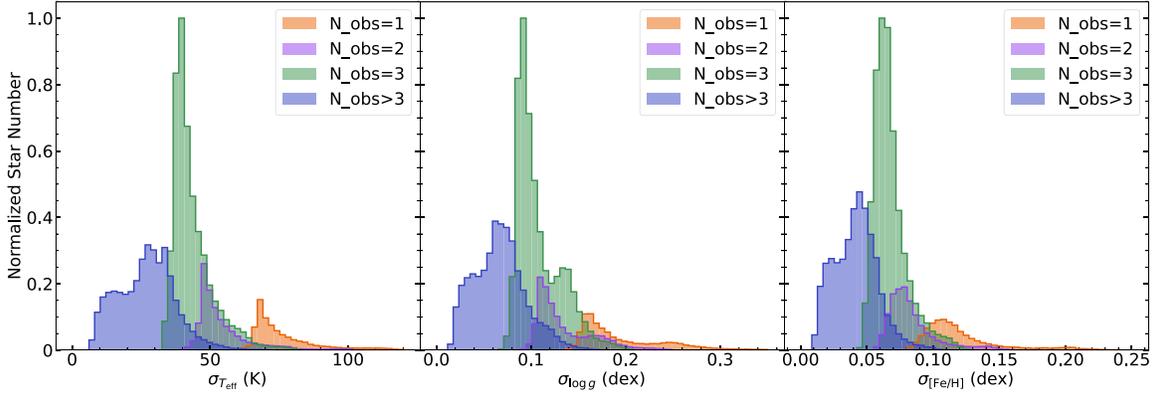

**Figure 13.** The distribution of uncertainties for the stellar parameters of $T_{\rm eff}$, $\log g$, and [Fe/H]. The number of visits are drawn with different colors.

**Table 3**
The Coefficients of the Multivariate Quadratic Polynomials

| Parameter | $a_0$ | $a_1$ | $a_2$ | $a_3$ | $a_4$ |
|---|---|---|---|---|---|
|  | $a_5$ | $a_6$ | $a_7$ | $a_8$ | $a_9$ |
| $T_{\rm eff}$ (K) | 1.415(−5) | 4.860(−4) | −1.469(−1) | −1.193(−1) | −2.661(−5) |
|  | 4.249(+2) | ⋯ | ⋯ | ⋯ | ⋯ |
| $\log g$ (dex) | 3.906(−8) | 4.010(−3) | 1.953(−6) | −4.172(−4) | −1.144(−2) |
|  | −8.810(−4) | −3.565(−6) | 5.516(−8) | −2.943(−5) | 1.318(0) |
| [Fe/H] (dex) | −2.028(−9) | −1.102(−3) | 5.423(−8) | 3.330(−2) | −5.213(−3) |
|  | −4.501(−7) | 5.874(−7) | −1.477(−7) | 2.194(−4) | −4.426(−2) |
| $v \sin i$ (km s$^{-1}$) | −1.233(−6) | −1.915(−3) | 7.868(−5) | 1.449(−2) | 7.440(−1) |
|  | −2.337(−2) | −7.645(−5) | 5.849(−7) | −6.519(−4) | −4.170(+1) |

**Note.** The numbers in the round brackets represent the powers of 10. For instance, 1.415(−5) means $1.415 \times 10^{-5}$.

random uncertainties. The difference in parameters between the adopted values and those from each visit for a star is denoted as

$$\Delta\theta \equiv |\theta_{\rm spec} - \theta_{\rm star}|; \quad (22)$$

here, $\theta_{\rm spec}$ represents the parameter value from a specific visit, and $\theta_{\rm star}$ is the adopted parameter value of the star. By fitting these parameter differences with a quadratic polynomial, we are able to estimate the random uncertainties. The value of a random uncertainty tends to decrease with the increasing S/N of a spectrum. Similar to our approach with the systematic uncertainties, we also consider that the random uncertainty of each stellar parameter not only is correlated with the parameter itself and S/N but also depends on $T_{\rm eff}$. Then, we apply multivariate quadratic polynomials for this estimation,

$$\begin{aligned}\Delta T_{\rm eff} = &\, a_0 T_{\rm eff}^2 + a_1 {\rm S/N}^2 + a_3 T_{\rm eff} \cdot {\rm S/N} \\ &+ a_4 T_{\rm eff} + a_5 {\rm S/N} + a_6\end{aligned} \quad (23)$$

$$\begin{aligned}\Delta \log g = &\, a_0 T_{\rm eff}^2 + a_1 \log g^2 + a_2 {\rm S/N}^2 \\ &+ a_3 T_{\rm eff} + a_4 \log g + a_5 {\rm S/N} \\ &+ a_6 T_{\rm eff} \cdot \log g + a_7 T_{\rm eff} \cdot {\rm S/N} \\ &+ a_8 \log g \cdot {\rm S/N} + a_9\end{aligned} \quad (24)$$

$$\begin{aligned}\Delta[\mathrm{Fe/H}] = &\, a_0 T_{\mathrm{eff}}^2 + a_1 [\mathrm{Fe/H}]^2 + a_2 \mathrm{S/N}^2 \\ &+ a_3 T_{\mathrm{eff}} + a_4 [\mathrm{Fe/H}] + a_5 \mathrm{S/N} \\ &+ a_6 T_{\mathrm{eff}} \cdot [\mathrm{Fe/H}] + a_7 T_{\mathrm{eff}} \cdot \mathrm{S/N} \\ &+ a_8 [\mathrm{Fe/H}] \cdot \mathrm{S/N} + a_9 \end{aligned} \quad (25)$$

$$\begin{aligned}\Delta v \sin i = &\, a_0 T_{\mathrm{eff}}^2 + a_1 v \sin i^2 + a_2 \mathrm{S/N}^2 \\ &+ a_3 T_{\mathrm{eff}} + a_4 v \sin i + a_5 \mathrm{S/N} \\ &+ a_6 T_{\mathrm{eff}} \cdot v \sin i + a_7 T_{\mathrm{eff}} \cdot \mathrm{S/N} \\ &+ a_8 v \sin i \cdot \mathrm{S/N} + a_9. \end{aligned} \quad (26)$$

As in the estimation of the systematic uncertainties, we group the stellar parameters and calculate the weighted mean value of difference within each group. We determine the systematic uncertainties using Equations (23)–(26), and the fitting coefficients are presented in Table 3.

### 4.5. Final Uncertainties

After estimating the stellar parameters and uncertainties of 3,126,343 spectra from LAMOST-MRS DR9, we use Equation (13) to calculate the stellar parameters of stars with multiple visits. In this equation, $\theta_j$ represents the stellar parameter of each spectrum of a star, $w_j = 1/\sigma_{\theta,j}^2$ is the weight, and $\sigma_{\theta,j}$ is the uncertainty of $\theta_j$. The uncertainties of a stellar parameter for a star are calculated using the law of propagation of uncertainty,

$$\sigma_\theta = \sqrt{\sum_j \left(\frac{\partial \theta}{\partial \theta_j}\right)^2 \sigma_{\theta,j}^2}; \quad (27)$$

the partial derivative $\partial \theta / \partial \theta_j$ can be derived by Equation (13). Consequently, we obtain the stellar parameters and uncertainties of 497,412 FGK type stars.

Figures 12 and 13 show the distributions of uncertainties in the stellar atmospheric parameters. It is evident that the spectra with higher S/Ns and the stars with more visits exhibit smaller uncertainties. For $T_{\mathrm{eff}}$, the uncertainty peaks around 75 K for the spectra with S/N > 10, and generally decreases following an exponential law when $\sigma_{T_{\mathrm{eff}}} > 75$ K. Figure 14 display a different uncertainty distribution of $\log g$ for dwarfs and giants. The uncertainty of $\log g$ for dwarfs has a peak value of ~0.17 dex, whereas, for giants, it is from 0.18 to 0.30 dex, with larger uncertainty for giants with lower $\log g$. There is no clear relation between the uncertainty of [Fe/H] and the S/N of spectra, and the peak value of the uncertainty at ~0.12 dex for all the spectra. This indicates that reliable [Fe/H] can be derived even for the spectra of S/N as low as 10. Due to the resolution of LAMOST-MRS spectra, we consider the $v \sin i$ value less than 40 km s$^{-1}$ is unreliable.

### 4.6. Flags

We assign 12 flags to indicate the possible problems in the stellar parameters of stars with multiple visits, and these flags are listed in Table 4. The first six flags represent whether the median absolute deviation (MAD) of $T_{\mathrm{eff}}$, $\log g$, [Fe/H], $v \sin i$, RV of blue, and red arm is larger than 3 times their respective uncertainty. The MAD for a data set $\{x_1, x_2, x_3 \cdots x_n\}$ is

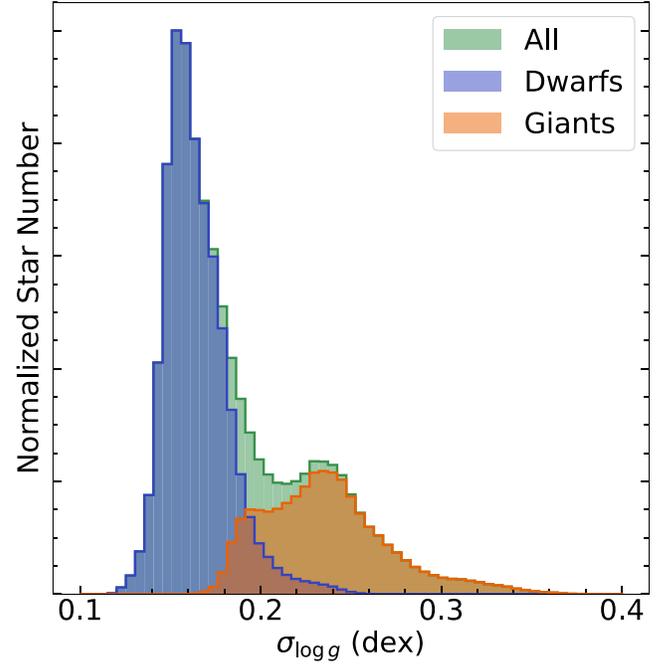

**Figure 14.** The distribution of uncertainties for $\log g$. It is obvious that the uncertainties of giants are larger than those of dwarfs.

**Table 4**
The Flags of the Stellar Parameters

| Flags | Comments |
| --- | --- |
| FLAG_MAD_TEFF | The MAD of $T_{\mathrm{eff}}$ is 3 times larger than the uncertainty. |
| FLAG_MAD_LOGG | Same as FLAG_MAD_TEFF but for $\log g$. |
| FLAG_MAD_FEH | Same as FLAG_MAD_TEFF but for [Fe/H]. |
| FLAG_MAD_VSINI | Same as FLAG_MAD_TEFF but for $v \sin i$. |
| FLAG_MAD_RV_B | Same as FLAG_MAD_TEFF but for blue arm spectral RV. |
| FLAG_MAD_RV_R | Same as FLAG_MAD_TEFF but for red arm spectral RV. |
| FLAG_DELTA_TEFF | The $\Delta T_{\mathrm{eff}}$ is 3 times larger than the uncertainty. |
| FLAG_DELTA_LOGG | Same as FLAG_DELTA_TEFF but for $\log g$. |
| FLAG_DELTA_FEH | Same as FLAG_DELTA_TEFF but for [Fe/H]. |
| FLAG_DELTA_VSINI | Same as FLAG_DELTA_TEFF but for $v \sin i$. |
| FLAG_DELTA_RV_B | Same as FLAG_DELTA_TEFF but for blue arm spectral RV. |
| FLAG_DELTA_RV_R | Same as FLAG_DELTA_TEFF but for red arm spectral RV. |

defined as

$$\mathrm{MAD} = \mathrm{median}(|x_i - \mathrm{median}(x_i)|). \quad (28)$$

The last six flags indicate whether the largest difference of these six stellar parameters for a star with multiple visits is larger than 3 times the uncertainty; the largest difference is defined as

$$\Delta \theta \equiv \theta_{\max} - \theta_{\min}; \quad (29)$$

here, $\theta_{\max}$ and $\theta_{\min}$ are the maximum and minimum values of parameter $\theta$, respectively.

For selecting the reliable $T_{\rm eff}$, $\log g$, [Fe/H], and $v \sin i$, we recommend choosing stars of zero for all 12 flags.

## 5. Summary

The massive collection of LAMOST spectra provides an unprecedented opportunity for studying the stellar formation and evolution. In this work, we describe our pipeline, LAMA, which is able to estimate the stellar parameters, including $T_{\rm eff}$, $\log g$, [Fe/H], RV, and $v \sin i$.

LAMA utilizes the CCF method to measure RV, and uses the template-matching method to estimate the stellar atmospheric parameters and $v \sin i$. The normalization is a necessary process, which impacts the accuracy of the stellar parameters determination. We use a third-order local regression fitting for the initial normalization. After measuring RV and estimating the initial stellar parameters, we calibrate the pseudocontinuum by dividing the nearest template and fitting the quotient. To estimate the accurate stellar parameters, we determine the resolution of LAMOST-MRS for each night and each fiber. LAMA estimates the stellar parameters with a $\chi^2$ minimization method, and the $\chi^2$ values calculated between the observed spectra and convolved templates.

By applying LAMA to the LAMSOT-MRS spectra of DR9, we derive the stellar parameters. We compare our results with those from reference high-resolution spectra of APOGEE, GALAH, and PASTEL. The comparison reveals a negligible bias, which demonstrates the accuracy of our result. We calibrate the stellar parameters using the common stars between our data and those from GALAH. The systematic and random uncertainties of our results are estimated, and the uncertainties for $T_{\rm eff}$ and [Fe/H] are around 75 K, 0.12 dex, respectively. For dwarfs stars, the uncertainty of $\log g$ is about 0.17 dex, while it is 0.18–0.30 dex for giants, and it decreases with increasing $\log g$.

In conclusion, we obtain the stellar parameters of 497,412 FGK type stars, corresponding to 3,126,343 spectra from LAMOST-MRS DR9. Our results will be used in the future research on the formation and evolution of stars and determination of elemental abundances.

## Acknowledgments

Our research is supported by the National Natural Science Foundation of China under grant Nos. 12090040, 12090044, 11833006, 12022304, 11973052, 11973042, U1931102, 12373036 and the National Key R&D Program of China No. 2019YFA0405502. We acknowledge the Scientific Instrument Developing Project of the Chinese Academy of Sciences, grant No. ZDKYYQ20220009, the science research grants from the China Manned Space Project with No. CMS-CSST-2021-B05 and the science research grants from the China Manned Space Project. This work is supported by Chinese Academy of Sciences President's International Fellowship Initiative grant No. 2020VMA0033 and Youth Innovation Promotion Association CAS (No. 202018); and. H.-L.Y. acknowledges the supports from Youth Innovation Promotion Association, Chinese Academy of Sciences. Guoshoujing Telescope (the Large Sky Area Multi-Object Fiber Spectroscopic Telescope, LAMOST) is a National Major Scientific Project built by the Chinese Academy of Sciences. Funding for the project has been provided by the National Development and Reform Commission. LAMOST is operated and managed by the National Astronomical Observatories, Chinese Academy of Sciences.

## Appendix A
## LAMA Stellar Parameters

Figure A1 shows the distributions of the calibrated stellar parameters.

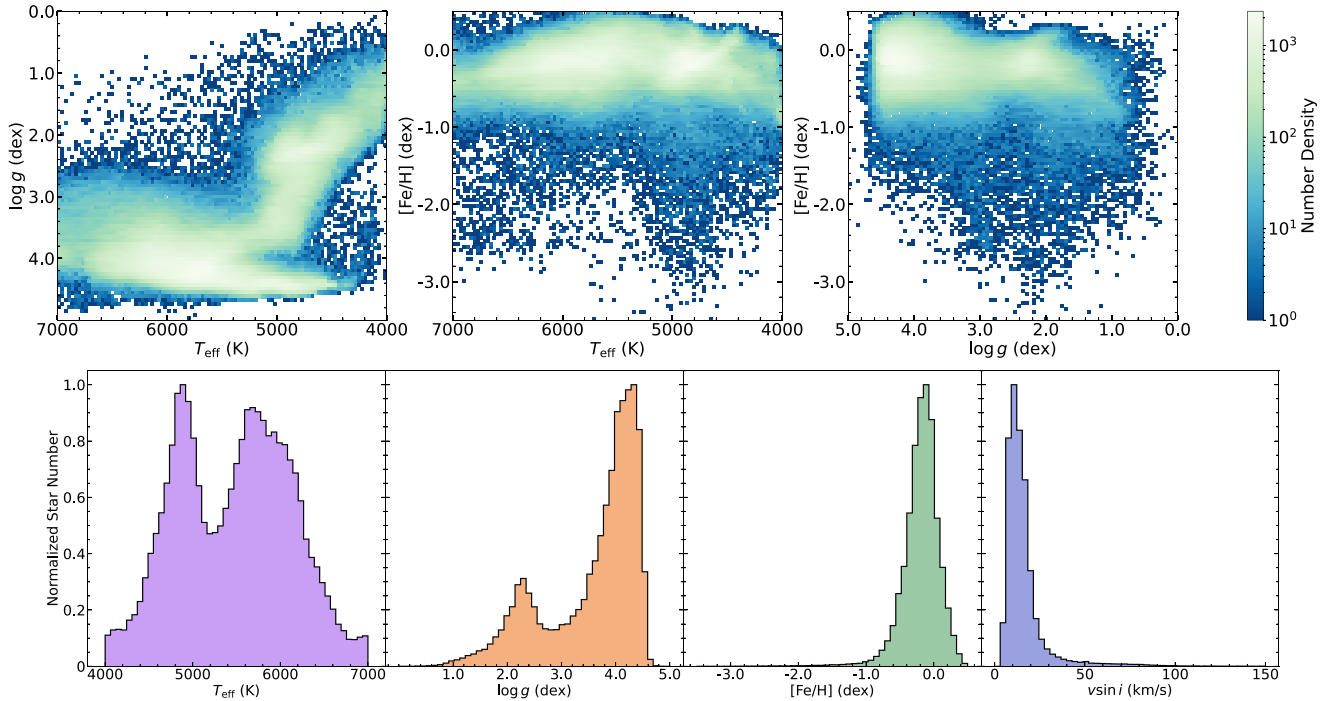

**Figure A1.** Upper panels: the left panel is the H-R diagram of calibrated LAMA stellar parameters, middle panel is [Fe/H] vs. $T_{\rm eff}$, and the right panel is [Fe/H] vs. $\log g$. Lower panels: the number of stars as functions of calibrated LAMA stellar parameter of $T_{\rm eff}$, $\log g$, [Fe/H], and $v \sin i$.

# Appendix B
# Comparison between LAMA and High-resolution Results

Figures B1, B2, and B3 show comparisons of stellar parameters between LAMA and APOGEE, GALAH, and PASTEL. Figures B4, B5, and B6 show the same, but after calibration.

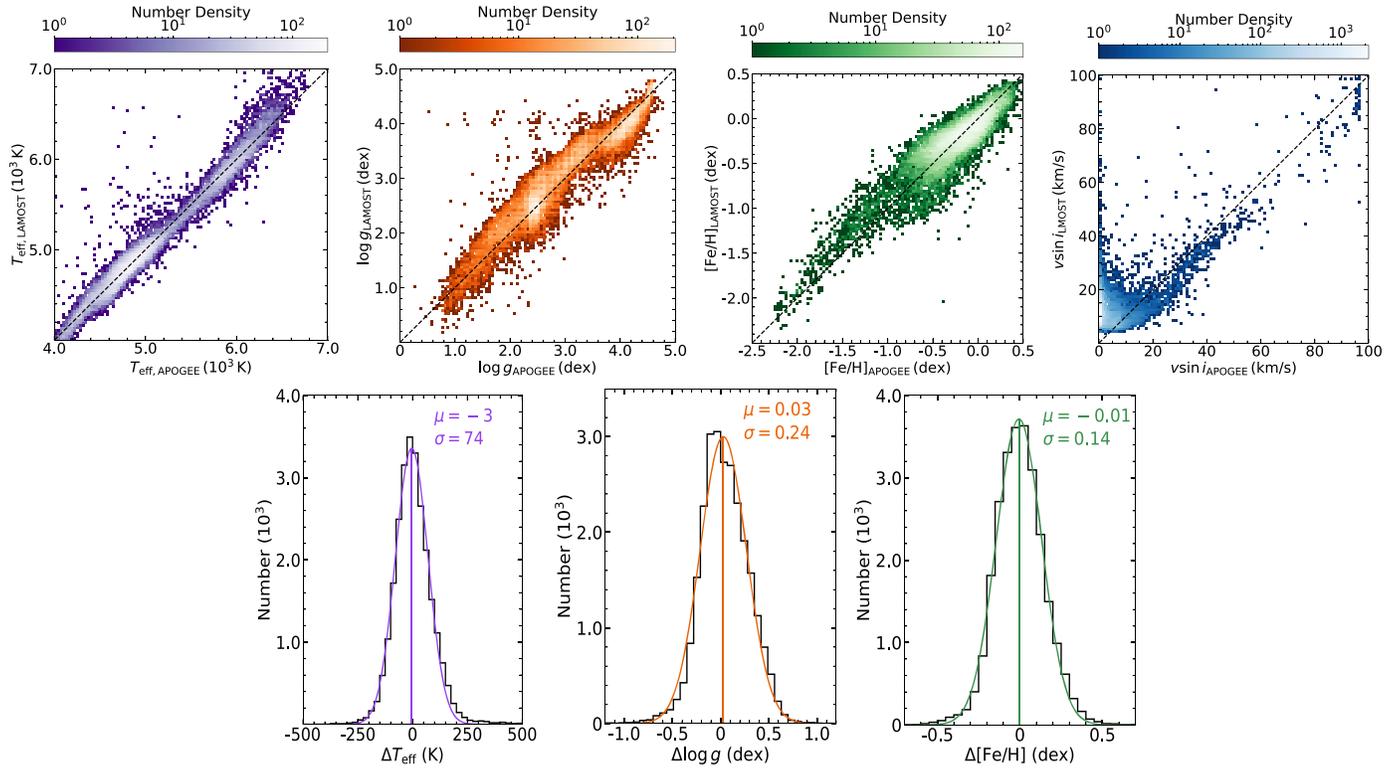

**Figure B1.** The comparisons of stellar parameters between LAMA and APOGEE. Upper panels: the relations between LAMA stellar parameters and APOGEE results. Lower panels: The number stars as functions of the differences between the LAMA stellar parameters and those from APOGEE, and the Gaussian profiles are the fitting results. The values of mean $\mu$ and variance $\sigma$ of Gaussian profiles are listed in the panels.

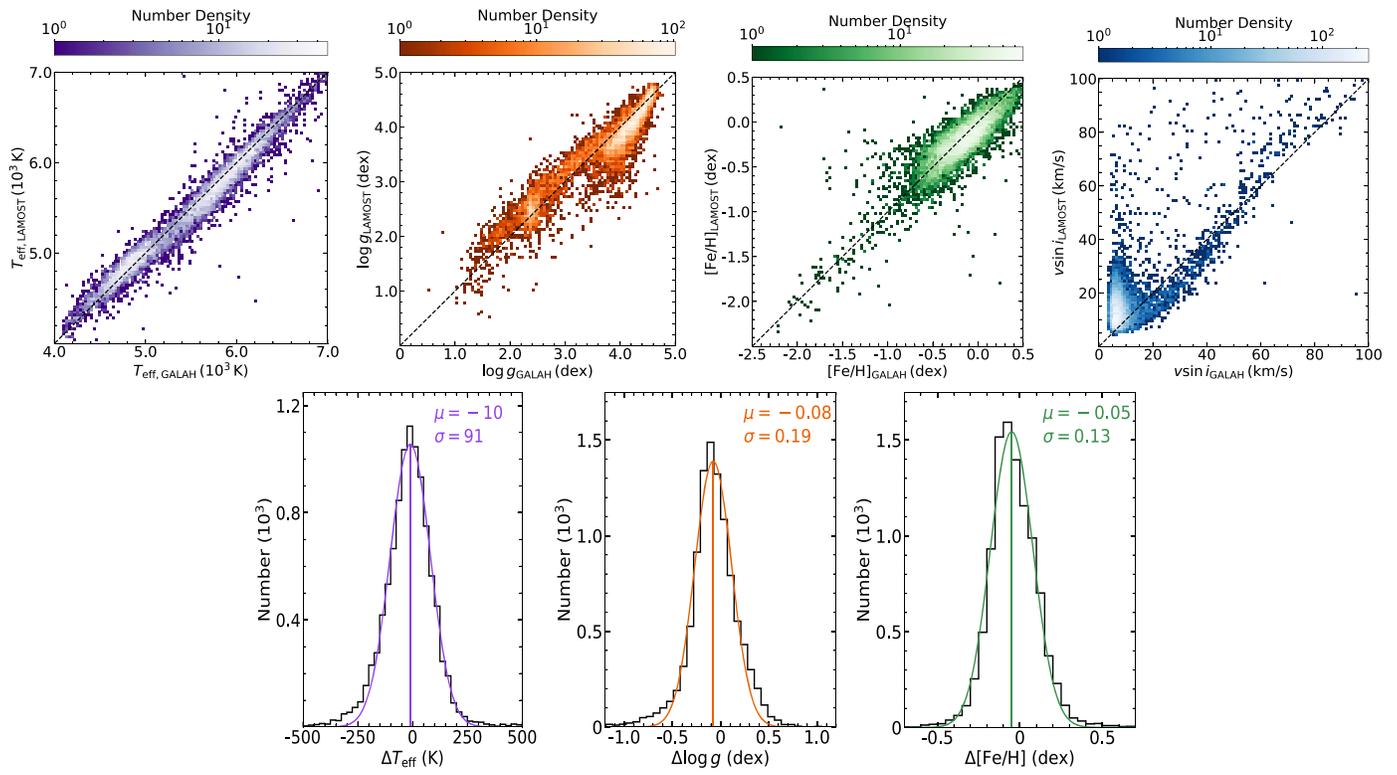

**Figure B2.** Same as Figure B1, but for GALAH.

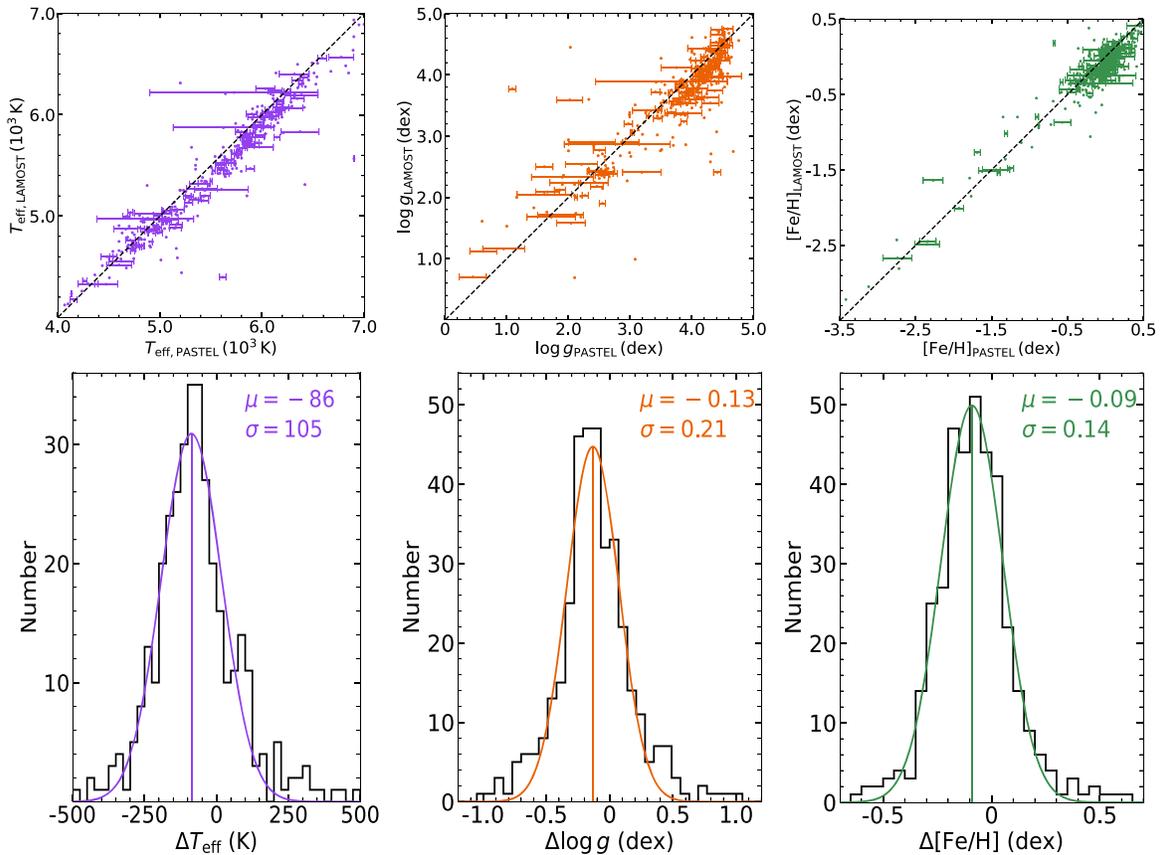

**Figure B3.** Same as Figure B1, but for PASTEL.

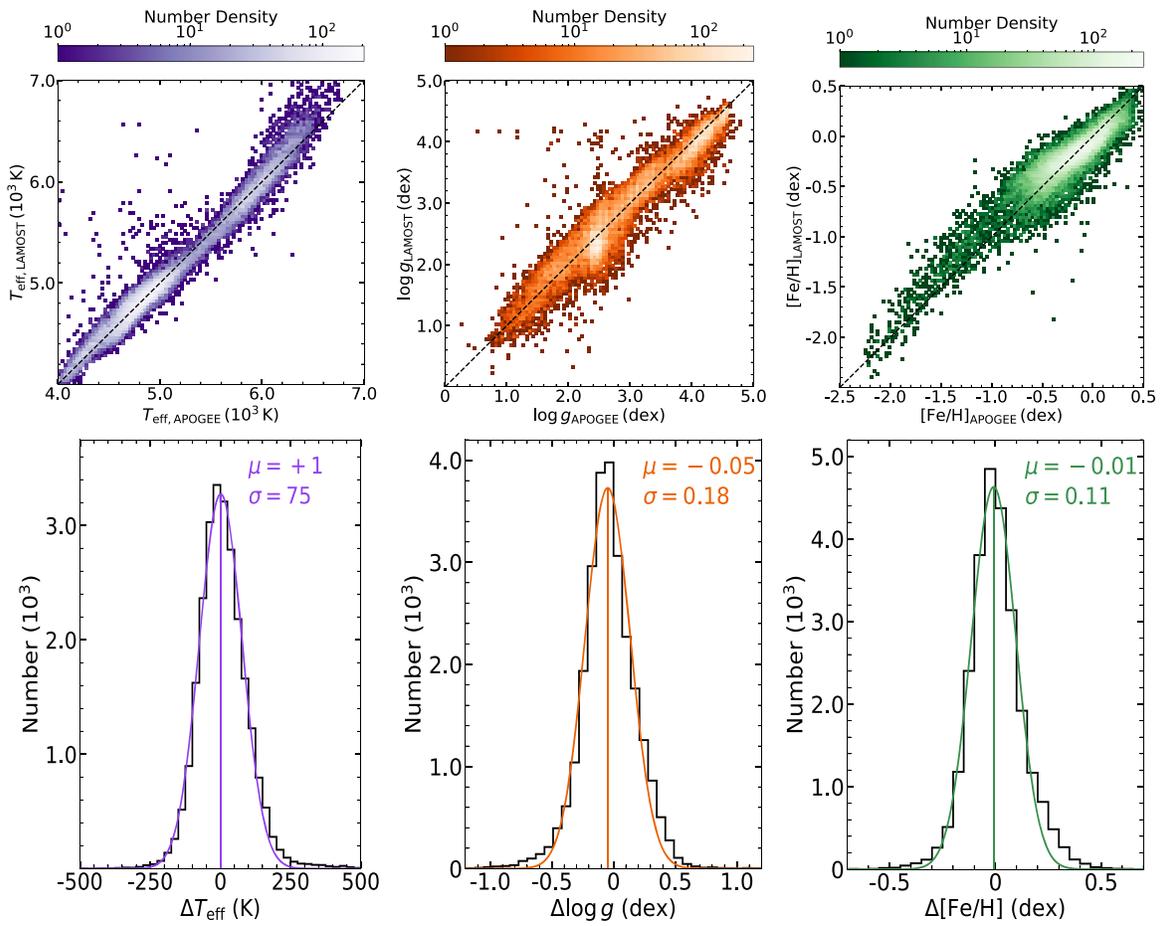

**Figure B4.** Same as Figure B1, but for the calibrated LAMA stellar parameters.

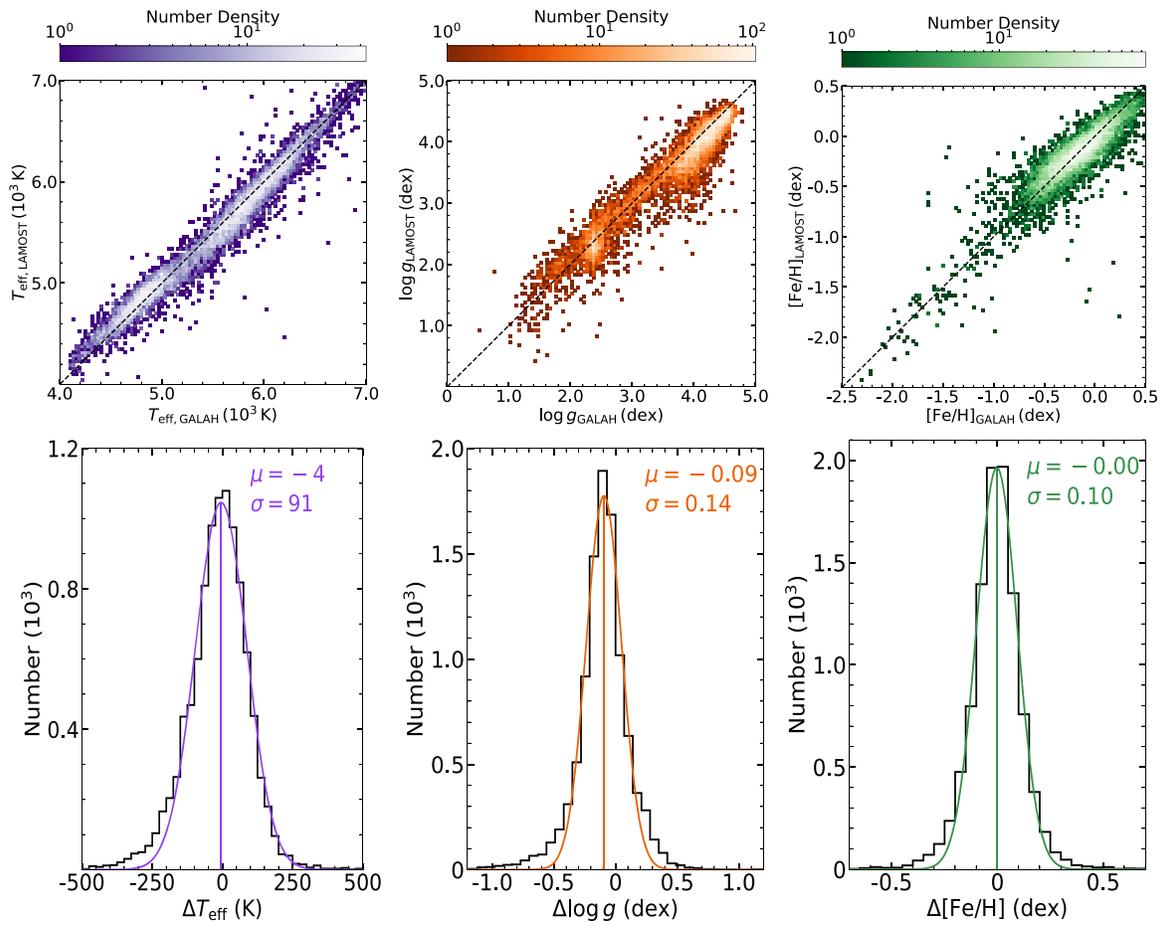

**Figure B5.** Same as Figure B2, but for the calibrated LAMA stellar parameters.

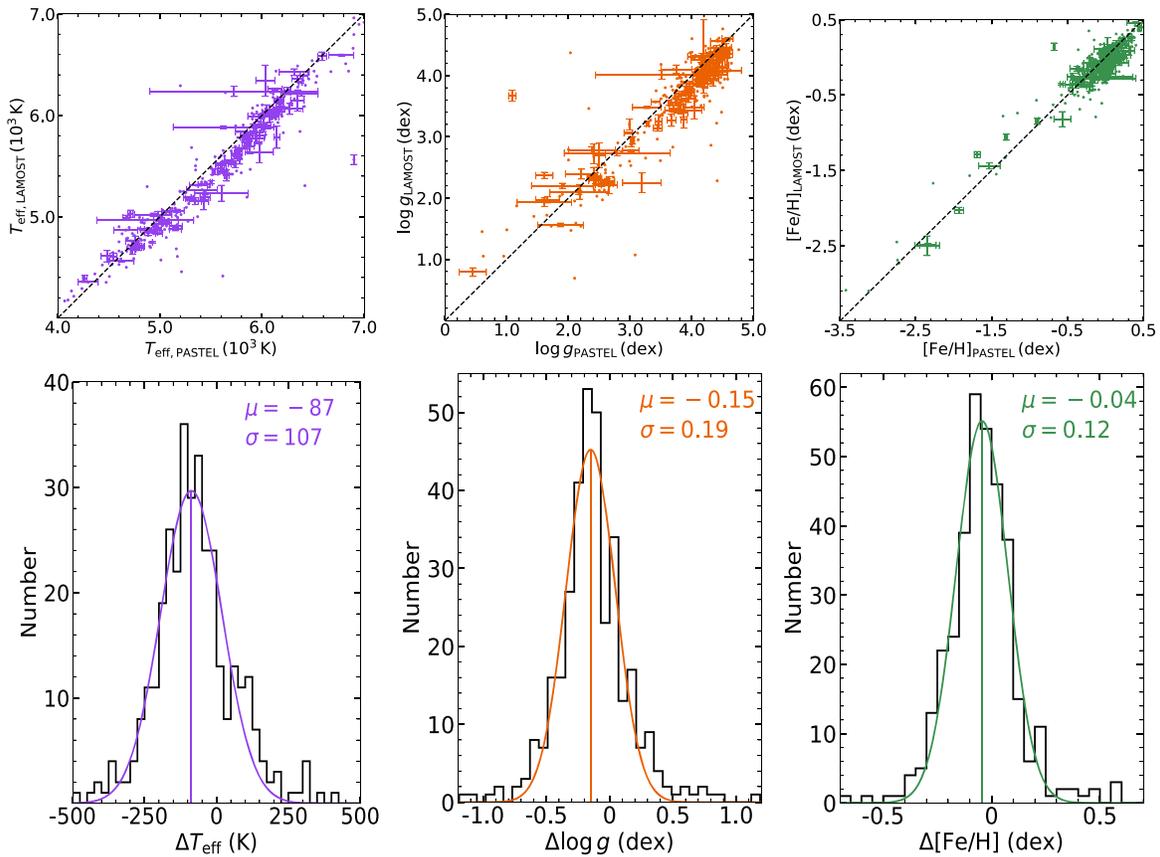

**Figure B6.** Same as Figure B3, but for the calibrated LAMA stellar parameters.

### ORCID iDs

Chun-qian Li ⓘ https://orcid.org/0000-0002-6647-3957
Jian-rong Shi ⓘ https://orcid.org/0000-0002-0349-7839
Hong-liang Yan ⓘ https://orcid.org/0000-0002-8609-3599
Zhong-rui Bai ⓘ https://orcid.org/0000-0003-3884-5693
Jiang-tao Wang ⓘ https://orcid.org/0000-0002-2316-8194
Ming-yi Ding ⓘ https://orcid.org/0000-0001-6898-7620